%% file: sample-sigconf.tex
\documentclass[sigconf]{acmart}

\usepackage{enumitem}
\usepackage{booktabs} 
\usepackage{subfigure}
\usepackage[linesnumbered,ruled,vlined]{algorithm2e}
\SetKwRepeat{Do}{do}{while}%
\SetKwFor{ForEach}{foreach}{do}{end}
\SetKwInput{KwData}{Input}
\SetKwInput{KwResult}{Output}

\usepackage{calrsfs}
\DeclareMathAlphabet{\pazocal}{OMS}{zplm}{m}{n}

\newcommand{\Db}{\pazocal{D}}
\newcommand{\PE}{\pazocal{PE}}

\setcopyright{rightsretained}

\acmConference[]{SIGIR 2017 eCom}{August 2017}{Tokyo, JAPAN} 
\acmYear{2017}
\copyrightyear{2017}

\begin{document}
\title{How to Balance Privacy and Money through Pricing Mechanism in Personal Data Market}

\author{Rachana Nget}
\affiliation{%
  \institution{Kyoto University}
  \city{Kyoto}
  \country{Japan}
}
\email{rachana.nget@db.soc.i.kyoto-u.ac.jp}

\author{Yang Cao}
\affiliation{%
  \institution{Emory University}
  \city{Atlanta} 
  \state{Georgia} 
  \country{USA}
}
\email{ycao31@emory.edu}

\author{Masatoshi Yoshikawa}
\affiliation{%
  \institution{Kyoto University}
  \city{Kyoto} 
  \country{Japan}}
\email{yoshikawa@i.kyoto-u.ac.jp}

\renewcommand{\shortauthors}{Rachana Nget, Yang Cao, Masatoshi Yoshikawa}
\renewcommand{\shorttitle}{How to Balance Privacy and Money through Pricing Mechanism}

\begin{abstract}

In the big data era, personal data is, recently, perceived as a new oil or currency in the digital world. Both public and private sectors wish to use such data for studies and businesses. However, access to such data is restricted due to privacy issues. Seeing the commercial opportunities in gaps between demand and supply, the notion of \textit{personal data market} is introduced. While there are several challenges associated with rendering such a market operational, we focus on two main technical challenges: (1) How should personal data be fairly traded under a similar e-commerce platform? (2) How much should personal data be worth in trade? 

In this paper, we propose a practical personal data trading framework that strikes a balance between money and privacy. To acquire insight on user preferences, we first conduct an online survey on human attitude toward privacy and interest in personal data trading. Second, we identify five key principles of the personal data trading central to designing a reasonable trading framework and pricing mechanism. Third, we propose a reasonable trading framework for personal data, which provides an overview of how data are traded. Fourth, we propose a balanced pricing mechanism that computes the query price and perturbed results for data buyers and compensation for data owners (whose data are used) as a function of their privacy loss. Finally, we conduct an experiment on our balanced pricing mechanism, and the result shows that our balanced pricing mechanism performs significantly better than the baseline mechanism.

\end{abstract}



\maketitle

\input{samplebody-conf}

\bibliographystyle{ACM-Reference-Format}
\bibliography{Mendeley} 

\end{document}

%% file: samplebody-conf.tex
\vspace{-0.7em}
\section{Introduction}

Personal data is, recently, perceived as a new \textit{oil} or \textit{currency} in the digital world. A massive volume of personal data is constantly produced and collected every second (i.e., via smart devices, search engines, sensors, social network services, etc.). These personal data are extraordinarily valuable for the public and private sector to improve their products or services. However, personal data reflect the unique value and identity of each individual; therefore, the access to personal data is highly restricted. For this reason, some large Internet companies and social network services provide free services in exchange for their users' personal data. Demand for personal data for research and business purposes excessively increases while there is practically no safe and efficient supply of personal data. Seeing the commercial opportunities rooted in gaps between demand and supply, the notion of \textit{personal data market} is introduced. This notion has transformed perceptions of personal data as an undisclosed type to a \textit{commodity}, as noted in \cite{ferretti2014eu} and \cite{Owns1996}. To perceive personal data as a commodity, many scholars, such as \cite{Ghosh2015}, \cite{Li2014}, \cite{Riederer2011}, and \cite{Roth2012}, have asserted that a monetary compensation should be given to real data producers/owners for their privacy loss whenever their data are accessed. Thus, personal data could be traded under the form of e-commerce where buying, selling, and financial transaction are done online.  However, this type of commodity might be associated with private attributes, so it should not be classified as one of the three conventional types of e-commerce goods (i.e., physical goods, digital goods, and services, as noted in \cite{joshi2017basic}). This privacy attribute introduces a number of challenges and requires different trading approach for this commodity called personal data. How much money should data buyers pay, and how much money should data owners require for their privacy loss from information derived from their personal data? One possible way is to assign the price in corresponding to the amount of privacy loss, but how to quantify privacy loss and how much money to be compensated for a metric of privacy loss are the radical challenges in this market.

\vspace{-0.7em}
\subsection{Personal Data Market}
The personal data market is a \textit{sound} platform for securing the personal data trading. What is traded as defined in \cite{Li2014} is a noisy version of statistical data. It is an aggregated query answer, derived from users' personal data, with some random noise included to guarantee the privacy of data owners. The injection of random noise is referred to as \textit{perturbation}. The magnitude of perturbation directly impacts the \textit{query price} and amount of data owners' \textit{privacy loss}. A higher query price typically yields a lower degree of perturbation (less noise injection).

In observing the published results of true statistical data, an adversary with some background knowledge (i.e., sex, birth date, zip code, etc.) on an individual in the dataset can perform linkage attacks to identify whether that person is included in the results. For instance, published anonymized medical encounter data were once matched with voter registration records (i.e., birth date, sex, zip code, etc.) to identify the medical records of the governor of Massachussetts, as explained in \cite{Dwork2014}. Therefore, statistical results should be subjected to perturbation prior to publication to guarantee an absence of data linkages.

\begin{figure}[h]
\setlength{\belowcaptionskip}{-10pt}
\setlength{\abovecaptionskip}{2pt}
  \includegraphics[width=\linewidth]{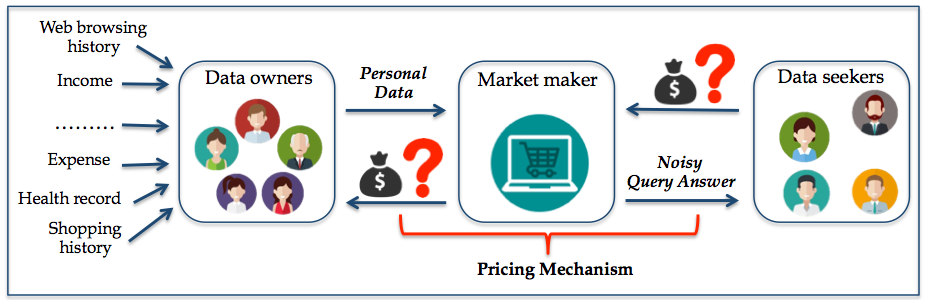}
  \caption{How much is personal data worth?}
  \label{fig:intro_pricing_mechanism}
\end{figure}

As is shown in Figure \ref{fig:intro_pricing_mechanism}, three main participants are involved: data owners, data seekers/buyers, and market maker. Data owners contribute their personal data and receive appropriate monetary compensation. Data buyers pay a certain amount of money to obtain their desirable noisy statistical data. Market maker is a trusted mediator between the two key players, as no direct trading occurs between two parties. A market maker is entrusted to compute a query answer, calculate query price for buyers and compensation for owners, and most importantly design a variety of payment schemes for owners to choose from.

\vspace{-0.6em}

The personal data market could be considered as the integration of Consumer-to-Business (C2B) and Business-to-Consumer (B2C) or Business-to-Business (B2B) e-commerce. On one side of the trading, the data owners as individuals provide their personal data to the market as is done in (C2B) e-commerce, though, at this point, no trading is done. On another end of the framework, the market maker sells statistical information to data buyers as an individual or company which is similar to (B2C) and (B2B) trading. This is when the trading transactions are completed in this framework. The study of such a market framework could initiate a new perception on the new forms of e-commerce.

The existence of personal data market will make abundance of personal data including sensitive but useful data safely available for various uses, giving rise to many sophisticated developments and innovations. For this reason, several start-up companies have developed online personal data trading sites and mobile applications following this market orientation. These sites are \textit{Personal}\footnote{www.personal.com}, and \textit{Datacoup}\footnote{www.datacoup.com}, which aim at creating personal data vaults. They buy the raw personal data from each data owner and compensate them accordingly. However, some data owners are not convinced to sell their raw data (without perturbation). For \textit{Datacoup}, payment is fixed at approximately \$8 for SNS and financial data (i.e., credit/debit card transactions). It is questionable whether \$8 is reasonable compensation, and how this price was decided. Another source of inefficiency is related to the absence of data buyers. This can create problems if buyers are not interested in such types of collected data. In addition, \textit{CitizenMe} and \textit{digi.me} recently launched personal data collection mobile applications that help data owners collect and store all of their personal data in their devices. Although the framework connects buyers to data owners, it might be inefficient and impractical for buyers to buy individual raw data one at a time. Moreover, as no pricing mechanism is offered, data owners and buyers must negotiate the prices on their own, which may not be efficient because not all data owners know or truthfully report the price of their data. This can result in an obstruction of trading operations. Based on lessons learned from such start-ups, we can conclude what they are missing is a \textit{well-designed trading framework}, that explains the principles of trading, and \textit{pricing mechanism}, that balances the money and privacy traded in the market.

To make this market operational, there are many challenges from all disciplines, but we narrow down fundamental technical challenges to two factors:

\vspace{-0.2em}
\begin{itemize}[leftmargin=5mm]
  \setlength{\parskip}{0pt}
  \setlength{\itemsep}{2pt plus 1pt}
  \item \textbf{Trading framework for personal data}: How should personal data be fairly traded? In other words, how should a \textit{reasonable trading framework be designed to respectively prevent circumvention from buyers on arbitrage pricing and from data owners on untruthful privacy valuation}? 
  \item \textbf{\textit{Balanced} pricing mechanism}: How much should personal data be worth? How should a price that balances data owners' privacy loss and buyers' payment be computed? This balance is crucial in \textit{convincing} data owners and data buyers to participate in the personal data market.
\end{itemize}
\vspace{-2mm}

\subsection{Contribution}
To address the above challenges more precisely, we first conducted a survey on human attitudes toward privacy and interest in personal data trading (Section \ref{sec:survey}). Second, from our survey analysis and from previous studies, we identify five key principles of personal data trading (Section \ref{subsec:principles}). Third, we propose a reasonable trading framework (Section \ref{subsec:framework}) that provides an overview of how data are traded and of transactions made before, during, and after trade occurs. Fourth, we propose a balanced pricing mechanism (Section \ref{sec:pricing_mechanism}) that computes the price of a noisy aggregated query answer and that calculates the amount of compensation given to each data owner (whose data are used) based on his or her actual privacy loss. The main goal is to balance the benefits and expenses of both data owners and buyers. This issue has not been addressed in previous researches. For instance, a theoretical pricing mechanism \cite{Li2014} has been designed in favor of data buyers only. Their mechanism empowers buyer to determine the privacy loss of data owners while assuming that data owners can accept an infinite privacy loss. Instead, our mechanism will empower both data owners and buyers to fully control their own benefits and expenses. Finally, we conduct an experiment on a survey dataset to simulate the results of our mechanism and prove the efficiency of our mechanism relative to a baseline pricing mechanism (Section \ref{sec:evaluation}).

\section{Survey Result} \label{sec:survey}
To develop deeper insight into personal data trading and to collect data for our experiment, we conducted an online survey delivered through a crowdsourcing platform. In total, 486 respondents from 46 different states throughout the USA took part in the survey. The respondents were aged 14 to older than 54 and had varying education backgrounds, occupations, and incomes. For our survey, respondents were required to answer 11 questions. Due to space limitations, We only discuss the more significant questions posed.

\textbf{Analysis 1:} For four types of personal data: \textit{Type 1} (commute type to school/work), \textit{Type 2} (yearly income), \textit{Type 3} (yearly expense on medical care), \textit{Type 4} (bank service you're using), the following results were obtained.

\vspace{-1em}
\begin{figure}[h]
\centering
\setlength{\belowcaptionskip}{-8pt}
\setlength{\abovecaptionskip}{0pt}
\mbox{\subfigure[Can sell Vs. Cannot sell.]{\includegraphics[width=1.63in]{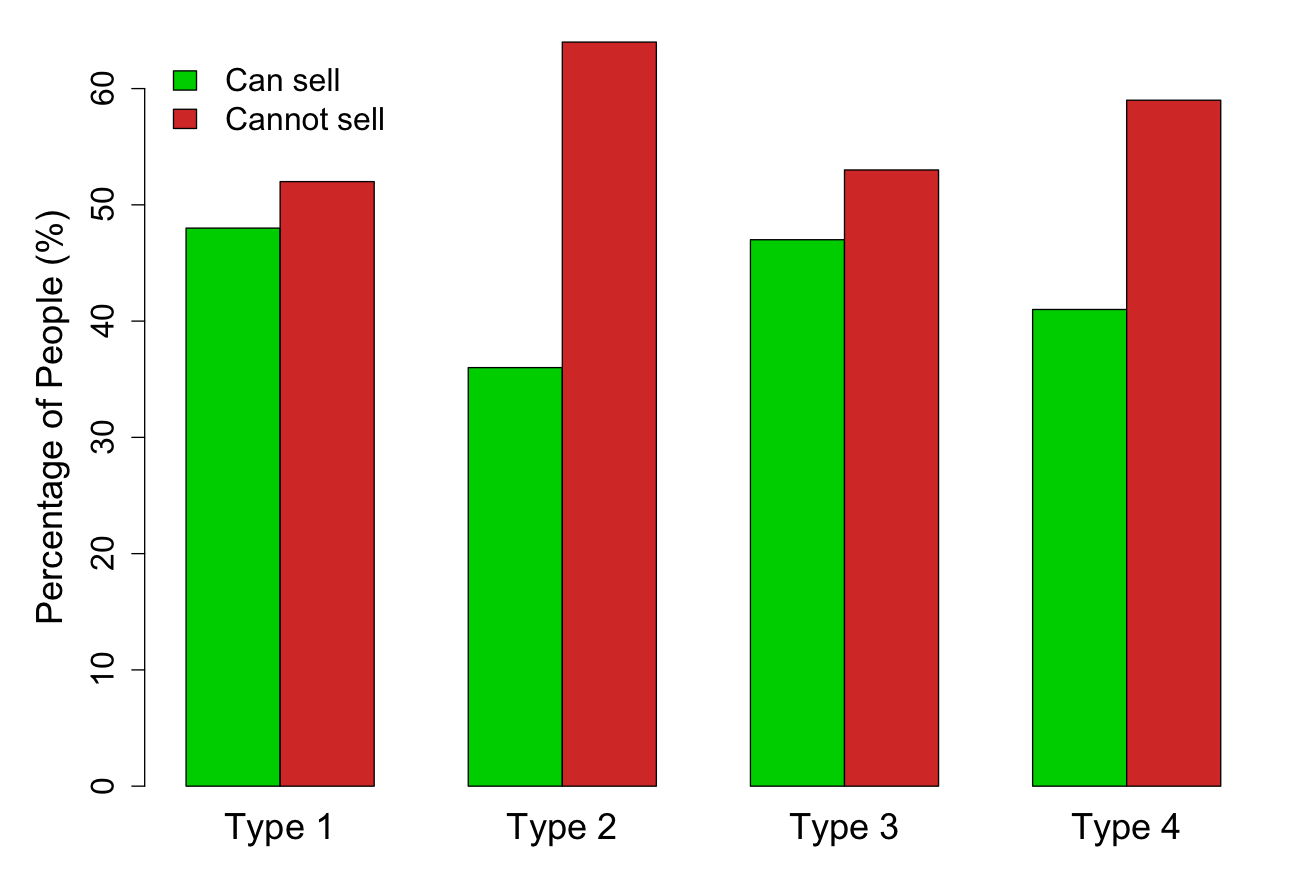}}\quad
\subfigure[How much to sell.]{\includegraphics[width=1.63in]{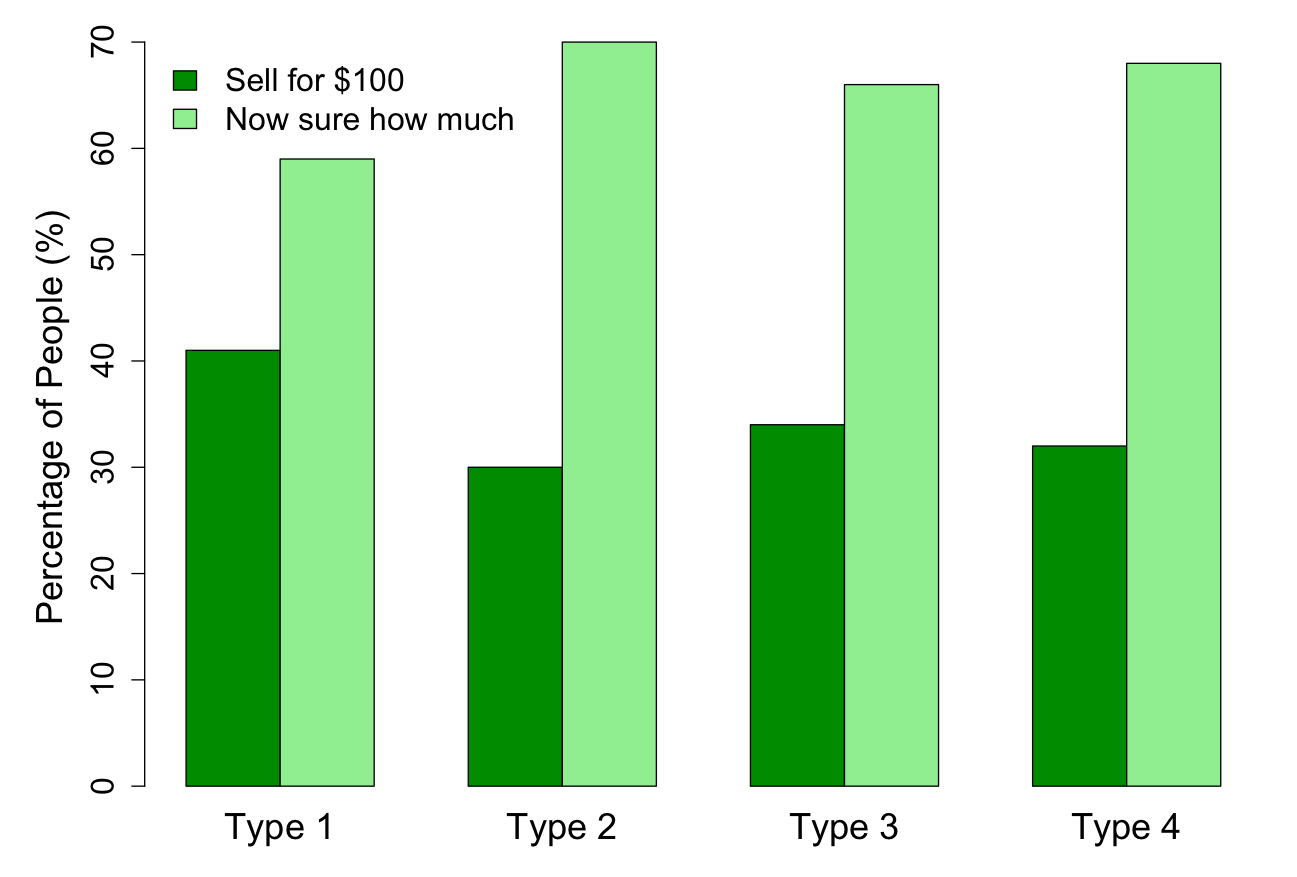} }}
\caption{Types of data to sell/not to sell.} \label{fig:type_to_sell}
\end{figure}

More than 50\% of the respondents said they \textit{cannot sell} the data (see Figure \ref{fig:type_to_sell}a), and more than 50\% of those who \textit{can sell} said that they \textit{do not know how much to sell} (see Figure \ref{fig:type_to_sell}b). 

Most of the participants stated that they do not know how much their data are worth, highlighting one of the above mentioned challenges related to the personal data market. Similarly, \cite{acquisti2013privacy} noted that it is very difficult for data owners to articulate the exact valuation of their data.  

\begin{figure}[h]
\centering
\setlength{\belowcaptionskip}{-9pt}
\setlength{\abovecaptionskip}{0pt}
\mbox{\subfigure[Interest in selling \textit{anonymized} data.]{\includegraphics[width=1.6in]{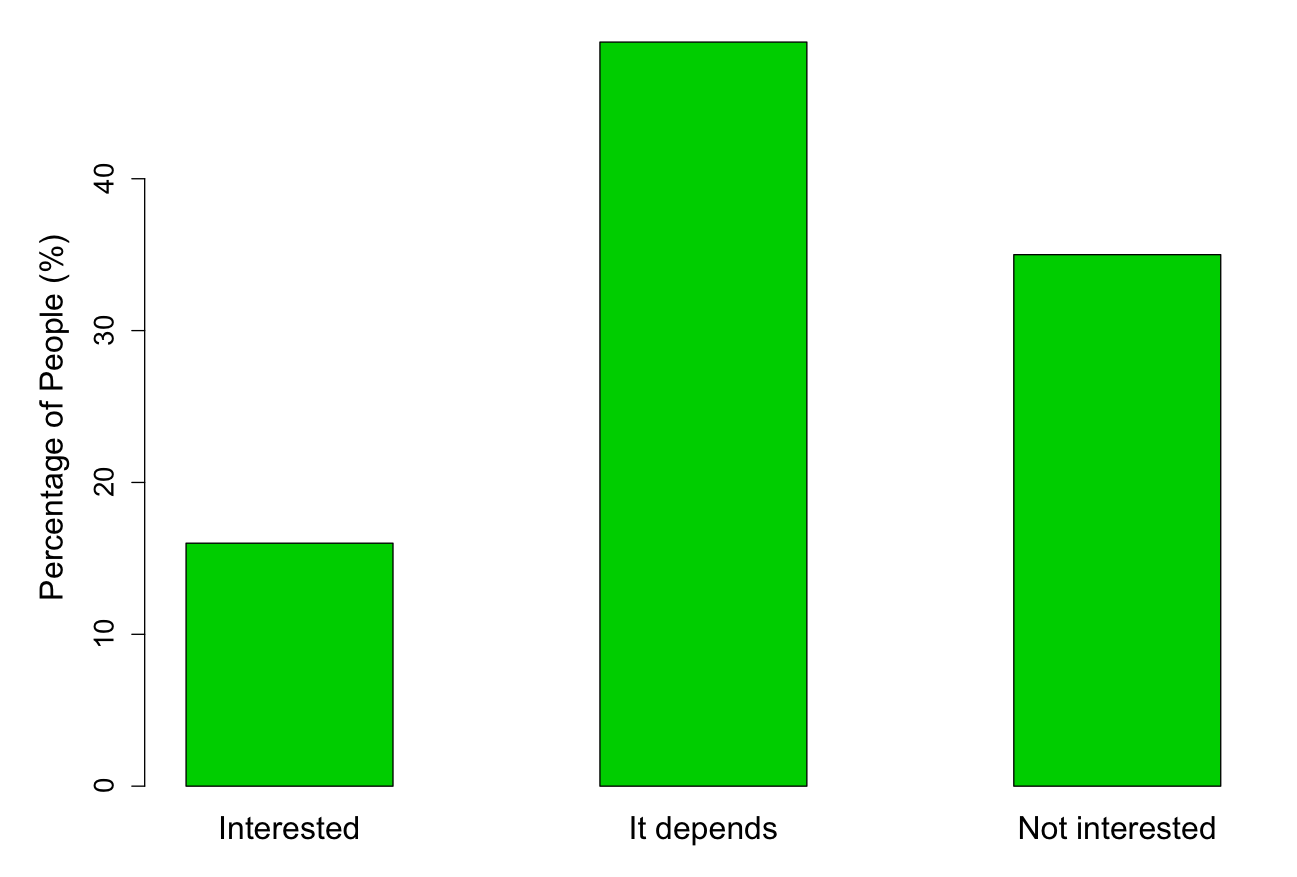}}\quad
\subfigure[Interest in selling both \textit{anonymized} and \textit{altered} data.]{\includegraphics[width=1.6in]{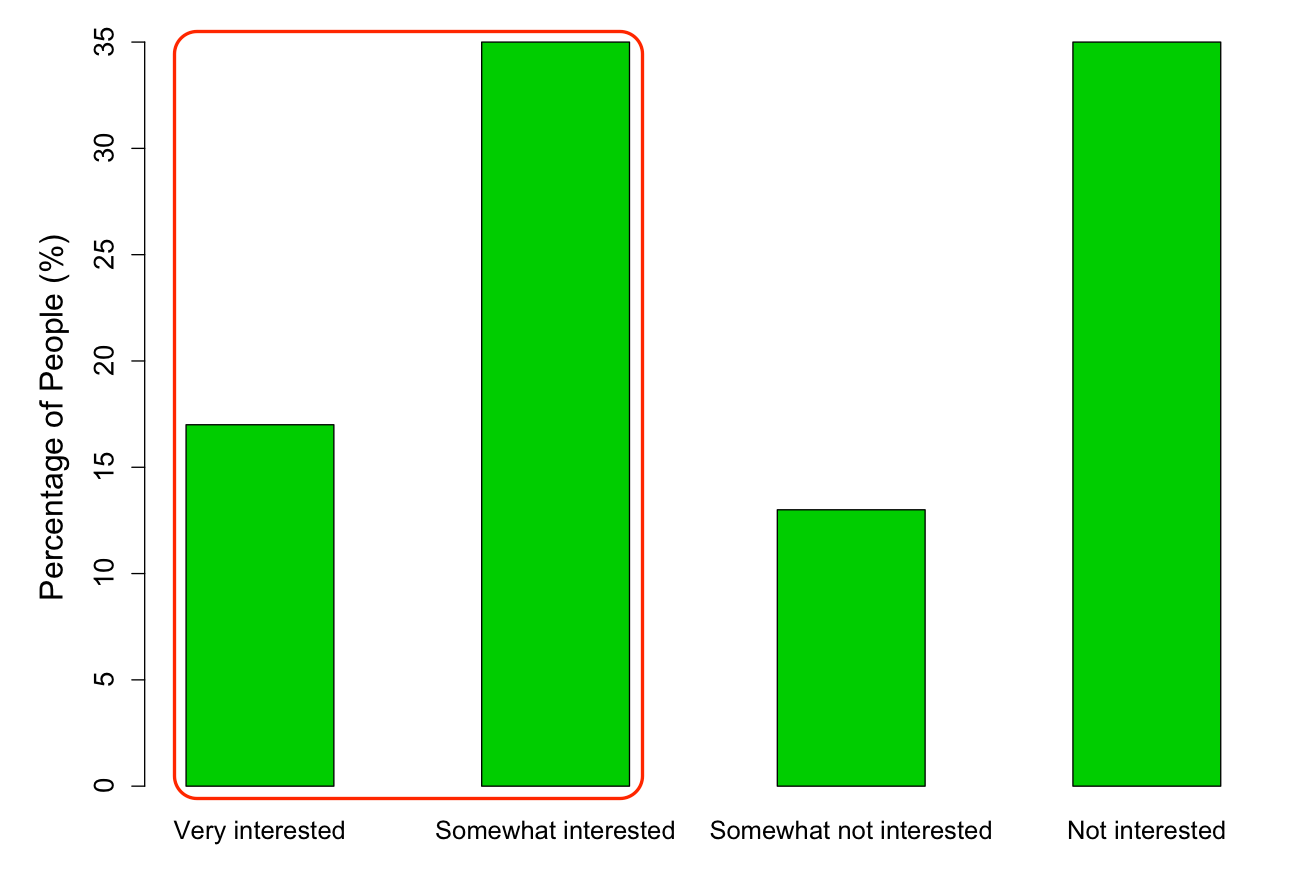} }}
\caption{Interest in selling personal data.} \label{fig:interest_in_selling}
\end{figure}

\textbf{Analysis 2:} When asked to sell their \textit{anonymized} personal data, 49\% of respondents said \textit{It depends on type of personal data and amount of money}, 35\% were \textit{Not interested}, and 16\% were \textit{Interested} (see Figure \ref{fig:interest_in_selling}a). However, if providing more privacy protection by both \textit{anonymizing} and \textit{altering} (\textit{perturbing}) real data, more than 50\% of the respondents became interested in selling, meaning that more people are now convinced to sell their data under such conditions. (see Figure \ref{fig:interest_in_selling}b).

\vspace{-1.6em}

\textit{Anonymization} does not convince people to sell their personal data. Providing extra privacy protection via data \textit{alteration} or \textit{perturbation} on the \textit{anonymized} data might make them feel more convinced and safer to sell their data.

\begin{figure}[h]
\centering
\setlength{\belowcaptionskip}{-10pt}
\setlength{\abovecaptionskip}{-2pt}
\mbox{\subfigure[\textit{Alteration levels} on data.]{\includegraphics[width=1.6in]{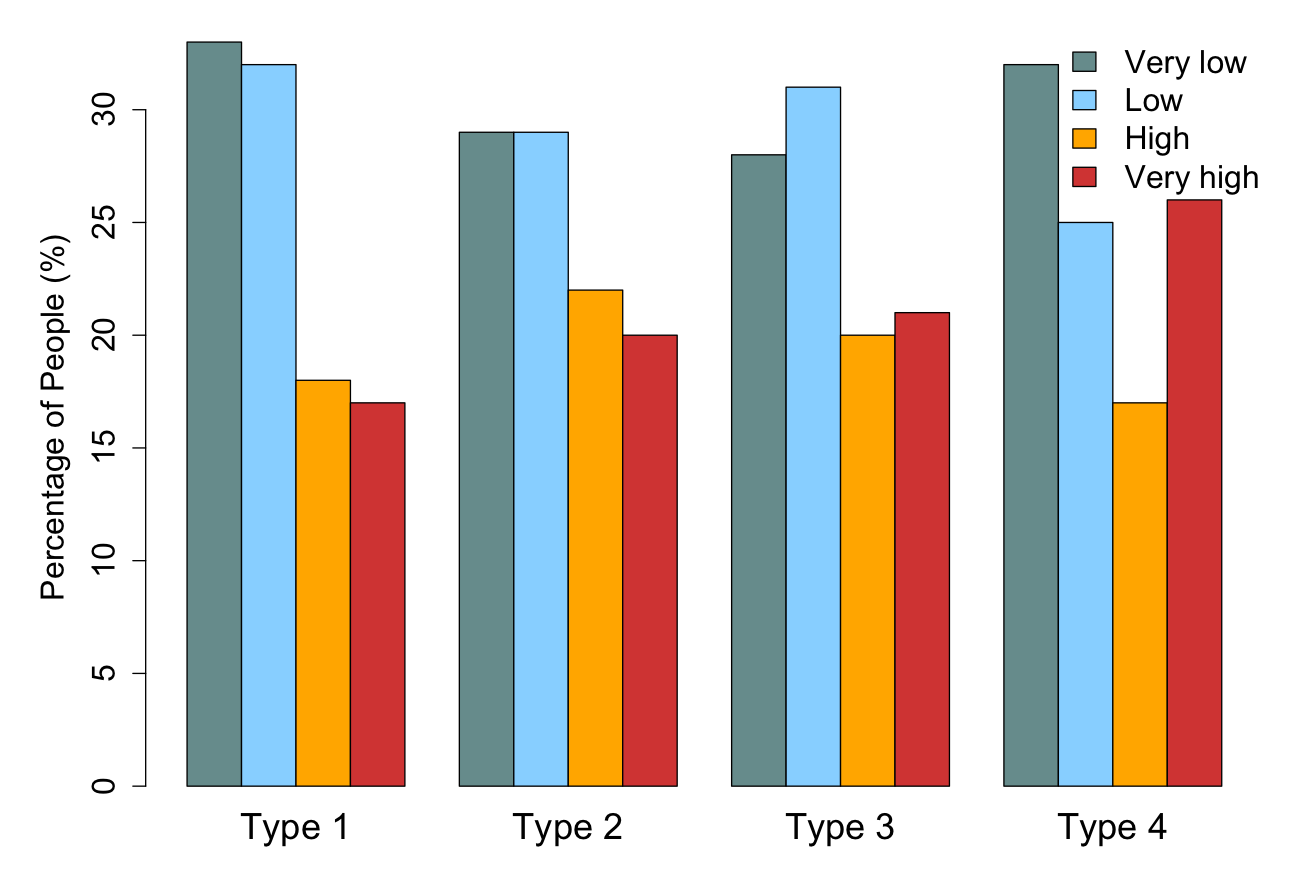}}\quad
\subfigure[Payment schemes.]{\includegraphics[width=1.6in]{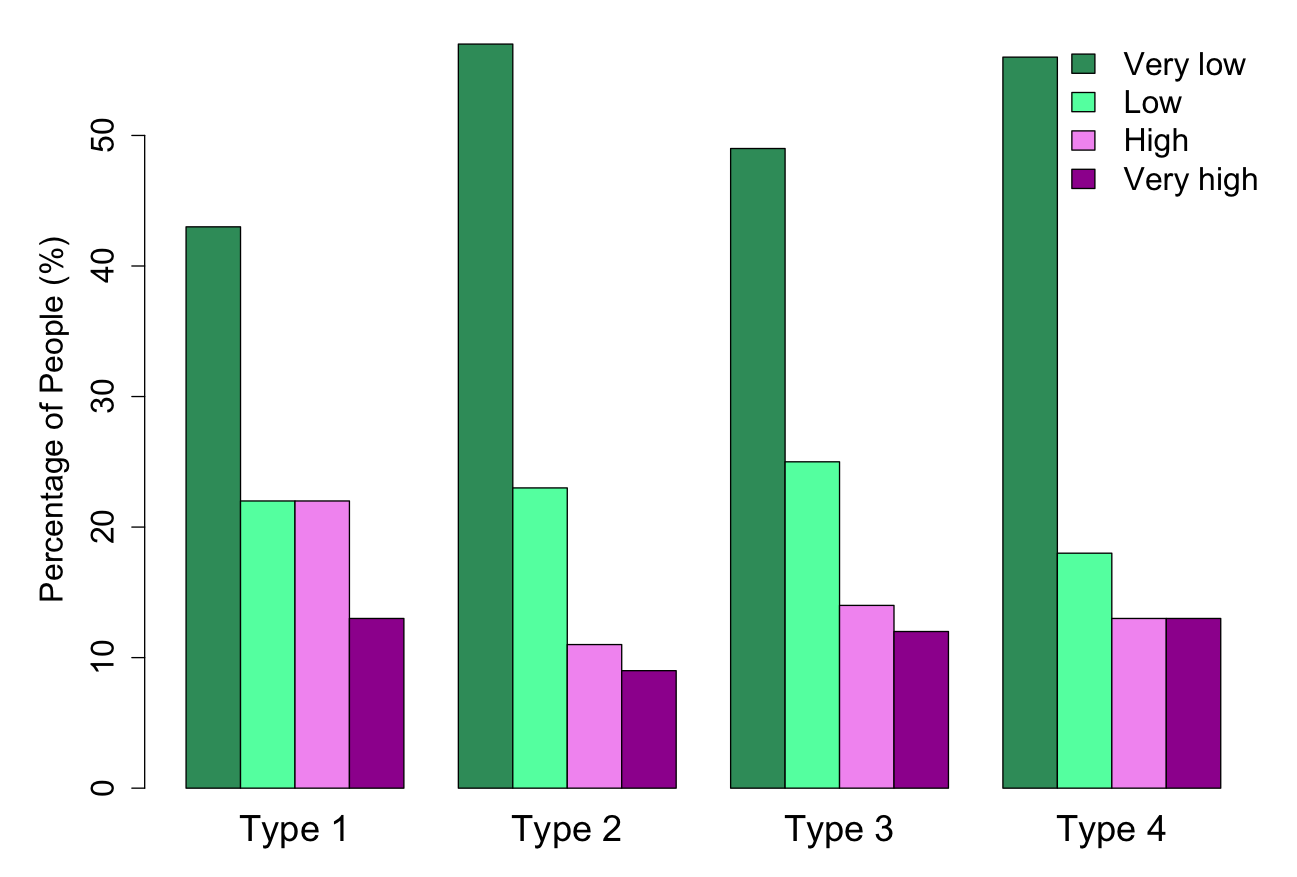}}}
\caption{Preferences in privacy and money.} \label{fig:preferences_in_selling}
\end{figure}

\textbf{Analysis 3:} With regard to alteration/perturbation, the respondents were asked to select their preferred privacy level: \{very low, low, high, very high\}, in other words, \textit{how much they want to alter/perturb their real data}. A \textit{very low} level of alteration (low noise injection) denotes a low privacy protection, but more monetary compensation. As a result (see Figure \ref{fig:preferences_in_selling}a), alteration levels were found to vary across the four types of data. Similarly, the preferred payment schemes (see Figure \ref{fig:preferences_in_selling}b) varied throughout all the data types. A human-centric study \cite{staiano2014money} also showed that people value different categories of data differently according to their behaviors and intentional levels of self-disclosure; as a result, location data are valued more highly than communication, app, and media data.  

\vspace{-1em}

Privacy protection levels and desired payment schemes varied in between the data considered and among the respondents. In practice, people harbor different attitudes toward privacy and money. Thus, it is crucial to allow a personalized privacy level and payment scheme for each individual.

\begin{figure}[h]
  \centering
  \setlength{\belowcaptionskip}{-8pt}
\setlength{\abovecaptionskip}{0pt}
  \includegraphics[width=1.65in]{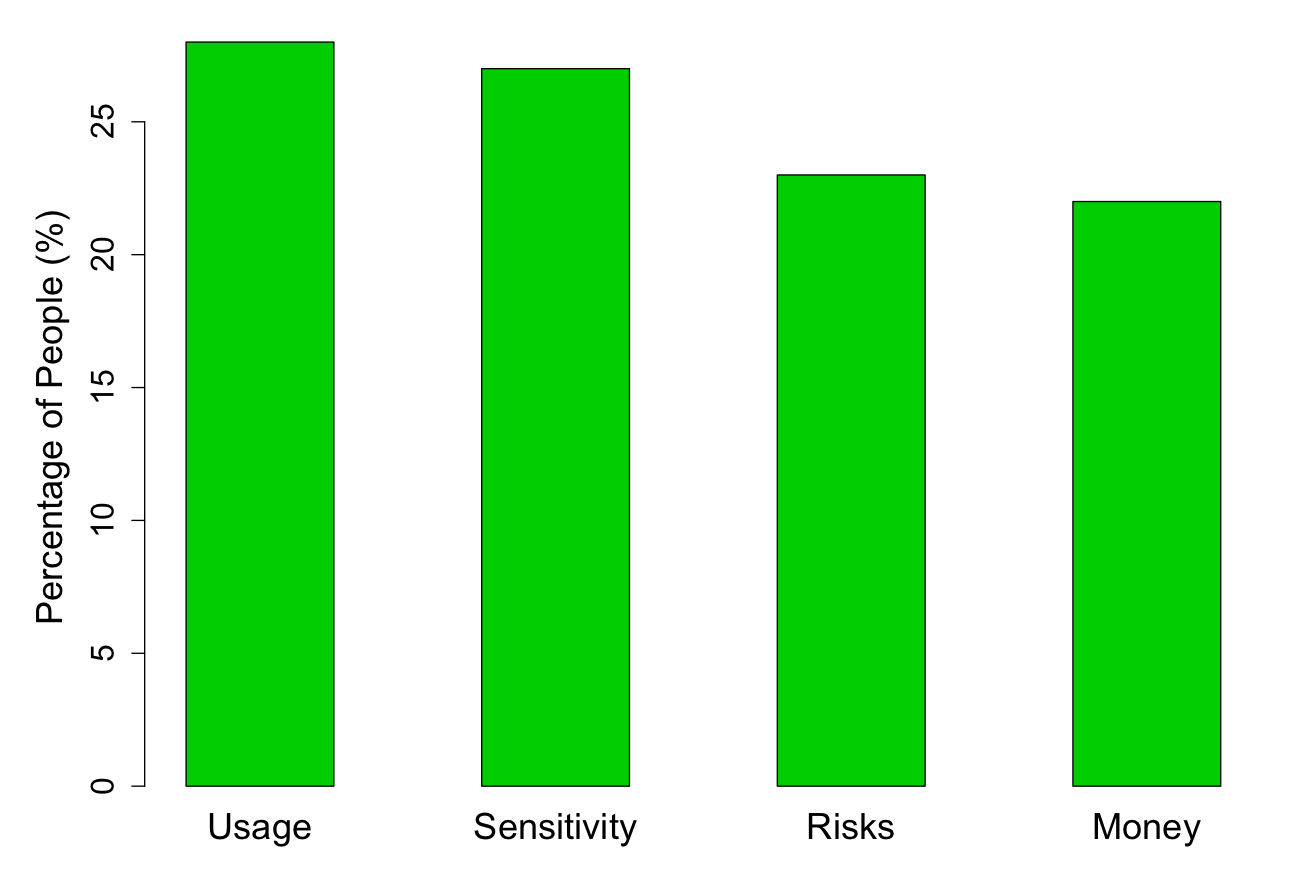}
  \caption{Importance of criteria when selling personal data.}
  \label{fig:importance_of_criteria}
\end{figure}

\textbf{Analysis 4:} Among the four given criteria to decide when selling personal data: \textit{usage} (who and how buyers will use your data), \textit{sensitivity} (sensitivity of data, i.e., salary, disease, etc.), \textit{risks} (future risks/impacts), and \textit{money} (to obtain as much money as possible),

In descending order, the participants valued the following: \textit{who and how the data will be used}, \textit{sensitivity}, \textit{future risks/impacts}, and \textit{money} (see Figure \ref{fig:importance_of_criteria}).

\vspace{-1em}

\textit{Money} is considered the least important criterion, while \textit{who and how data will be used} is considered the most important one when deciding to sell personal data. This implies that \textit{money cannot buy everything when the seller does not want to sell}.

\section{Trading Framework} \label{sec:framework}

All notations used in this study are summarized in Table \ref{tab:notation}.

\vspace{-0.3em}
\begin{table}
\setlength{\belowcaptionskip}{-10pt}
\setlength{\abovecaptionskip}{0pt}
  \caption{Summary of notations.}
  \label{tab:notation}
  \begin{tabular}{|c||l|}
	\hline
     Notation & Description\\
	\hline
    $ u_{i}, b_{j}$ & Data owner $i$, Data buyer $j$\\
    $x_{i}$  & Data element of $u_{i}$\\
    $\hat{\varepsilon_{i}}$  & Maximum tolerable privacy loss of $u_{i}$\\
    $w_{i}$  & Payment scheme of $u_{i}$\\
    $\varepsilon_{i}$  & Actual privacy loss of $u_{i}$ in query computation\\
    $w_{i}(\varepsilon_{i})$  & Compensation to $u_{i}$ for losing $\varepsilon_{i}$\\
    $x$  & Dataset consisting of a data element of all $u_{i}$\\
    $Q$  & Linear aggregated query requested by the buyer\\
    $W_{max}$  & Maximum budget of the buyer\\
    $W_{p}, W_{r}$  & Query price, Remaining budget of the buyer\\
	$Q(x)$  & True query answer\\
	$P(Q(x))$  & Perturbed query answer (with noise)\\
	$RMSE$  & Root mean squared error\\
    $\chi$  & Market maker's profit\\
    $W_{ab}$  & Available budget for query computation\\
    $RS$  & A representative sample of dataset $x$\\
    $h$  & Number of representative samples $RS$\\
    $\Phi$  & Number of perturbation run times\\
 
  \hline
\end{tabular}
\end{table}

\subsection{Key Principles of the Trading Framework}\label{subsec:principles}

To design a reasonable trading framework and a balanced pricing mechanism, it is important to determine the chief principles of the personal data trading framework. These key principles are derived from previous studies and from the four key analyses of our survey. The principles are categorized into five different groups: personalized differential privacy as a privacy protection, applicable query type, arbitrage-free pricing model, truthful privacy valuation, and unbiased result. To guarantee the data owner's privacy, personalized differential privacy injects some randomness into the result based on the preferred privacy level. It is also used as a metric to quantify the privacy loss of each data owner. With this personalized differential privacy guarantee, only some certain linear aggregated query types are applicable in this trading framework. Regarding pricing, a pricing model should be arbitrage-free and must not allow any circumventions on the query price from any savvy buyers. Similarly, such a framework should be designed to encourage data owners' truthful privacy valuation by providing them the right pricing scheme so that they will not benefit from any untruthful valuation. Finally, it is important to ensure the generation of unbiased/less biased query result without increasing query price, so a careful sample selection method is crucial.

\subsection*{A. Personalized Differential Privacy as a Privacy Protection} \label{sec:principle_differential_privacy}

The pricing mechanism should be capable of preserving data owner's privacy from any undesirable privacy leakages. To ensure privacy, differential privacy \cite{Dwork2014} plays an essential role in guaranteeing that the adversary could learn nothing about an individual while learning useful information about the whole population/dataset from observing the query result (despite some background knowledge about that individual). Given a privacy parameter $\varepsilon$, any private mechanisms (i.e., Laplace mechanism, Exponential mechanism, etc.) satisfy the $\varepsilon$-differential privacy level if the same result is likely to occur regardless of the presence or absence of any individual in the dataset as a result of random noise addition. A smaller $\varepsilon$ offers better privacy protection but is less accurate, resulting in a tradeoff between privacy and result accuracy. In our framework, we define $\varepsilon$ as the quantification of privacy loss of data owner as $\varepsilon$ and money are correlated.

\begin{definition}[$\varepsilon$-Differential Privacy \cite{Dwork2014}]\label{def:dp}
A random algorithm $ M: D \rightarrow R $ satisfies $ \varepsilon$-Differential Privacy ($\varepsilon$-DP) if the neighboring dataset $ x $, $ y $ $ \in $ $ D $ where $ D $ is a whole dataset and dataset $ x $ and $ y $ differs by only one record, and any set of $ S \subseteq Range (M) $,
\begin{equation}
\Pr(M(x) \in S) \leq exp( \varepsilon ) * \Pr(M(y) \in S)
\end{equation}
\end{definition}

In regard to differential privacy (DP), privacy protection is for the tuple level, which means that all users included in the dataset have the same privacy protection/loss $\varepsilon$ value (one for all). However, in practice, individuals may have different privacy attitude, as illustrated in our survey result, so allowing privacy personalization is considered critical, especially in the trading setting. We thus adopt the \textit{personalized differential privacy (PDP)} theory by \cite{Jorgensen2015}, which is derived from the above differential privacy. Each user can personalize his or her maximum tolerable privacy level/loss $\hat{\varepsilon_{i}}$, so any private mechanisms that satisfy $\hat{\varepsilon_{i}}$-differential privacy must guarantee each user's privacy up to their $\hat{\varepsilon_{i}}$. Users may set $\hat{\varepsilon_{i}}$ according to their privacy attitude with the assumption that $\hat{\varepsilon_{i}}$ is public and is not correlated with the sensitivity of data. This theory thus allows users' privacy personalization while offering more utility to data buyers.

\begin{definition}[Personalized Differential Privacy \cite{Jorgensen2015}]
Regarding the maximum tolerable privacy loss ${\hat{\varepsilon}}$ of each user and a universe of users $ U $, a randomized mechanism $ M: D \rightarrow R $  satisfies ${\hat{\varepsilon}}$-Personalized Differential Privacy (or ${\hat{\varepsilon}}$-PDP), if for every pair of neighboring datasets $ x, y \in D $ where $ x $ and $ y $ differs in data for user $ i $, and for any set of $ S \subseteq Range (M) $,
\begin{equation}
\Pr(M(x) \in S) \leq exp(\hat{\varepsilon}) * \Pr(M(y) \in S)
\end{equation}
\end{definition}

Both DP and PDP are theories, so a private mechanism is employed to realize these theories. \cite{Jorgensen2015} introduced two PDP private mechanisms: \textit{sampling} and \textit{exponential-like mechanisms}. Given a privacy threshold, the sampling mechanism samples a subset drawn from the dataset and then runs one of the private mechanisms (i.e., Laplace mechanism, etc.). The exponential-like mechanism, given a set of $\hat{\varepsilon}$, computes a score (probability) for each potential element in the output domain. This score is inversely related to the number of changes made in a dataset $x$ required for a potential value to become the true answer.

\begin{definition}[Score Function \cite{Jorgensen2015}] Given a function $f: \Db \rightarrow R$ and outputs $r \in Range(f)$ with a probability proportional to that of the exponential mechanism differential privacy \cite{Dwork2014}, $s(D, r)$ is a real-valued score function. The higher the score, the better $r$ is relative to $f(D)$. Assuming that $D$ and $D'$ differ only in the value of a tuple, denoted as $D \oplus D'$, 
\vspace{-0.3em}
\begin{equation}
s(D, r) = \underset{f(D')=r}{\mathrm{max}} -|D \oplus D'|
\end{equation}

\end{definition}

In PDP, each record or data owner has their own privacy setting $\hat{\varepsilon_{i}}$, so it is important to distinguish between different $D'$ that make a specific value to become the output. To formalize this mechanism, \cite{Jorgensen2015} defined it as follows.

\begin{definition}[$\PE$ Mechanism \cite{Jorgensen2015}] \label{def:PE_mechanism}
Given a function $f:D\rightarrow R$, an arbitrary input dataset $D \subset \Db$, and a privacy specification $\phi$, the mechanism $\PE_{\phi}^f (D)$ outputs $r\in R$ with probability

\begin{equation}\label{eq:PE_mechanism}
Pr[\PE_{\phi}^f (D) = r] = \dfrac{exp(\dfrac{1}{2}d_{f}(D,r,\phi))}{\sum_{q\in R}exp(\dfrac{1}{2}d_{f}(D,r,\phi))}
\end{equation}
where $d_{f}(D,r,\phi) = \underset{f(D')=r}{max} \sum_{i\in D \oplus D'} - \phi^{i_{u}}$
\end{definition}

In our framework, $\phi$ refers to a set of maximum tolerable privacy loss $\hat{\varepsilon_{i}}$ of all data owners in the dataset $x$. We apply this $\PE$ mechanism to guarantee that each data owner's privacy is protected despite data owners having different privacy requirements. The proof of this mechanism can be found in \cite{Jorgensen2015}.

\subsection*{B. Applicable Query Type} \label{sec:principle_query_type}
With background knowledge, the adversary may engage in linkage attacks on the published query answer and may eventually identify an individual from this answer. Therefore, any queries answered in this trading framework should guarantee that results do not reveal whether or not an individual is answering the query. DP or PDP can prevent the data linkage attacks on the published results of statistical/linear aggregated queries by introducing randomness. For these reasons, only statistical/linear aggregated queries should be allowed in the trading framework when the privacy is guaranteed by DP or PDP. \cite{Li2014} also adopted this query type in their proposed theoretical framework. 

\begin{definition}[Linear Query \cite{Li2014}]
Linear Query is a vector with real value $q = (q_{1},q_{2},...,q_{n})$. The computation of this query $q$ on a fixed-size data vector $x$ is the result of a vector product $q.x = q_{1}.x_{1}+...+q_{n}.x_{n}$.
\end{definition}

\subsection*{C. Arbitrage-free Pricing Model} \label{sec:principle_arbitrage_free}
\textit{Arbitrage-free} is a requisite property used to combat the circumvention of a savvy data buyer on the query price. For instance, a perturbed query answer with a larger $\varepsilon_{1}=1$ costs \$10 and that with a smaller $\varepsilon_{2}=0.1$ costs \$0.1. If a savvy buyer seeks a perturbed query answer with $\varepsilon=1$, he or she will buy the query answer with $\varepsilon_{2}=0.1$ 10 times to compute the average of them for the same result as $\varepsilon_{1}=1$ because $\varepsilon$ increases as the number of computation times $n$ increases $\varepsilon =$ ($n * \varepsilon_{2}$). This case is explained based on \textit{composition theorems} in \cite{Dwork2014}. Therefore, the buyer will never have to pay \$10 for the same result as the average of several cheap queries costing him/her only \$1. In \cite{Li2014}, the \textit{arbitrage-free} property is defined as follows:

\begin{definition}[Arbitrage-free \cite{Li2014}]
A pricing function $\pi(Q)$ is arbitrage-free if for every multiset $S={Q_{1},...,\\Q_{m}}$ and $Q$ can be determined from $S$, denoted as $S\rightarrow Q$, then: 

\begin{equation}
\pi(Q) \leq \sum_{i=1}^{m} \pi(Q_{i})
\end{equation}
\end{definition}

An explanation and discussion of query determinacy $(S\rightarrow Q)$ can be found in \cite{Li2014}.

\textit{Arbitrage-free pricing function}:
\cite{Li2014} proved that a pricing function $\pi(Q)$ can be made equal to the sum of all payments made to data owners if the framework is balanced. A framework is balanced if: \textbf{(1)} the pricing function $\pi$ and payment function to data owners are \textit{arbitrage-free}, and \textbf{(2)} the query price is cost-recovering, which means that the query price should not be less than that needed to compensate all data owners. In our framework, we simply adopt their \textit{arbitrage-free} property by ensuring that the query price $W_{q}$ is always greater than the compensation given to all data owners (whose data are accessed) for their actual privacy loss $\varepsilon_{i}$.

For simplicity, a buyer shall not be able to request the same query more than once because each data owner has his or her own $\hat{\varepsilon_{i}}$, so we must guarantee that their privacy loss is no greater than their specified $\hat{\varepsilon_{i}}$. Alternatively, market maker can predefine the sets of queries that buyer can ask for so that they can study relationships between all queries in advance to prevent arbitrage problems from emerging. However, this also limits the choice of query buyers can request, so our framework allows buyers to ask any linear aggregated queries but only once per query.

\subsection*{D. Truthful Privacy Valuation}\label{sec:principle_truthful_privacy_valuation}

Untruthful privacy valuation is an undesirable property leading to the generation of unacceptably high query prices. Without carefully designed payment schemes, some savvy data owners will always attempt to select any schemes that provide them more benefits, so they may intentionally report an unreasonably high privacy valuation. For instance, \cite{Li2014} applied a linear payment scheme ($w_{i} = c_{i}*\varepsilon$) and allowed each data owner to define the $c_{i}$. With the same $\varepsilon$, most data owners will always set very high $c_{i}$ values to maximize benefits.   

To encourage truthful privacy valuation, all data owners shall be provided with the suitable payment scheme corresponding to their privacy/risk attitudes so that untruthful valuations do not increase their benefits, as illustrated \cite{Aperjis2012}.

\vspace{-1.5em}
\begin{figure}[!h]
\setlength{\belowcaptionskip}{-10pt}
\setlength{\abovecaptionskip}{0pt}
   \includegraphics[width=0.8\linewidth]{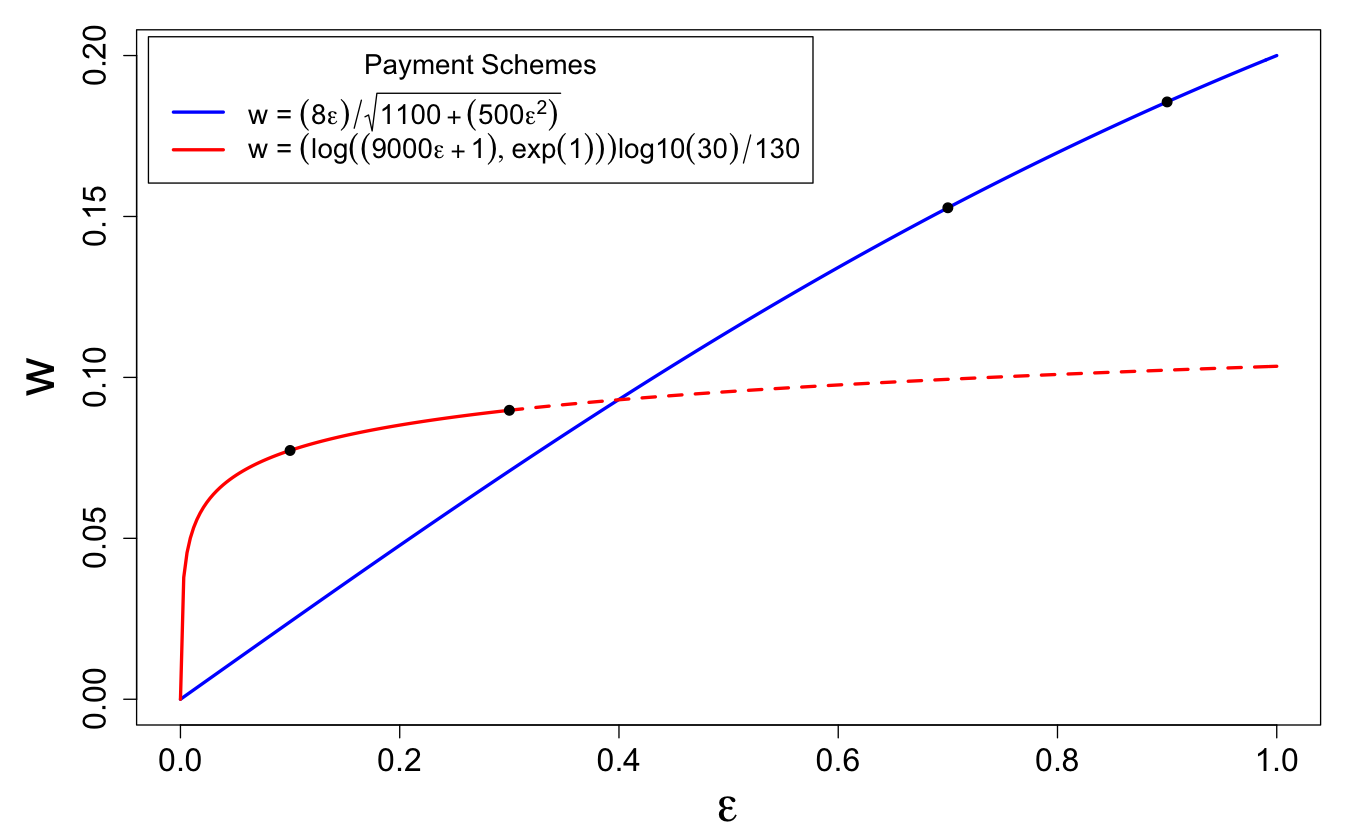}
  \caption{Payment Schemes.}
  \label{fig:payment_contract_function}
\end{figure}

\vspace{-10pt}
\begin{proposition}[Payment Scheme]
A payment scheme is a non-decreasing function $w:\varepsilon\rightarrow R^+$ representing a promise between a market maker and a data owner on how much data owner should be compensated for their actual privacy loss $\varepsilon_{i}$. Any non-decreasing functions can be denoted as payment schemes. For instance,

\vspace{-\topsep}
\begin{itemize}
  \setlength{\parskip}{0pt}
  \setlength{\itemsep}{1pt plus 1pt}
  \item {\textbf{Type A:}} This Logarithm function is designed to favor conservative (low-risk, low-return) data owners whose $\hat{\varepsilon}$ is small.
  	\begin{equation}\label{eq:contract_function_type_A}
  		w = \dfrac{log(30)*ln(9000\varepsilon + 1)}{130}
  	\end{equation}
  \item {\textbf{Type B:}} This Sublinear function is designed to favor liberal (high-risk, high-return) data owners whose $\hat{\varepsilon}$ is large.
  \vspace{-\topsep}
  	\begin{equation}\label{eq:contract_function_type_B}
  		w = \dfrac{8\varepsilon}{\sqrt{1100+500\varepsilon^2}}
  	\end{equation}
 \end{itemize}
 \vspace{-\topsep}

\end{proposition}

For our framework, we designed two different types of payment schemes, as illustrated in Figure \ref{fig:payment_contract_function}. The data owner shall select a payment scheme based on his or her privacy $\hat{\varepsilon}$ or risk orientation. Therefore, there is no reason for data owners to untruthfully report their privacy valuation $\hat{\varepsilon}$ because doing so would not provide them with any benefits. The market maker designs a pricing scheme, and the guidelines of a design should mainly depend on equilibrium theory of the supply and demand. In the present study, we only consider two types of functions to provide different options for conservative and liberal data owners. We will develop a more sophisticated scheme in our future work.

\subsection*{E. Unbiased Result}\label{sec:principle_unbiased_result}

Besides ensuring privacy protection and price optimization, unbiased result has been a crucial factor in trading. Buyers do not want to obtain a result that is biased or that is significantly different from the true result, so it is important to ensure the generation of an unbiased result. 

In our setting, we guarantee the generation of an unbiased/less biased result by randomly selecting data owners, among which both liberal and conservative data owners are equally likely to be selected. Employing the PDP assumption, data owner's $\hat{\varepsilon_{i}}$ value is not correlated with the sensitivity of data, so random selection best guarantees a less biased result.

Moreover, to optimize the query price, it is necessary to select a representative sample from a dataset because paying each individual data owner in the dataset (as in \cite{Li2014}) leads to the generation of very high query prices for the same level of data utility. Thus, sampling a good representative subset is very useful. We apply statistical sampling method to compute the number of data owners required for each representative sample given a dataset. A similar concept is employed in \cite{Aperjis2012}.

A personal data trading framework should adopt these five key principles to avoid certain issues and to obtain more optimal results. However, a similar study by \cite{Li2014} did not consider all of these key principles. First, data owners cannot personalize their privacy levels as they are assumed to accept infinite losses when more money is paid. Moreover, their mechanism cannot efficiently reduce query prices because a query is computed on the entire dataset, and data owners can easily untruthfully report their privacy valuation to maximize the amount of payment given a linear payment scheme.

\subsection{Personal Data Trading Framework}\label{subsec:framework}

\begin{figure}[!h]
\centering
\setlength{\belowcaptionskip}{-10pt}
\setlength{\abovecaptionskip}{2pt}
   \includegraphics[width=\linewidth]{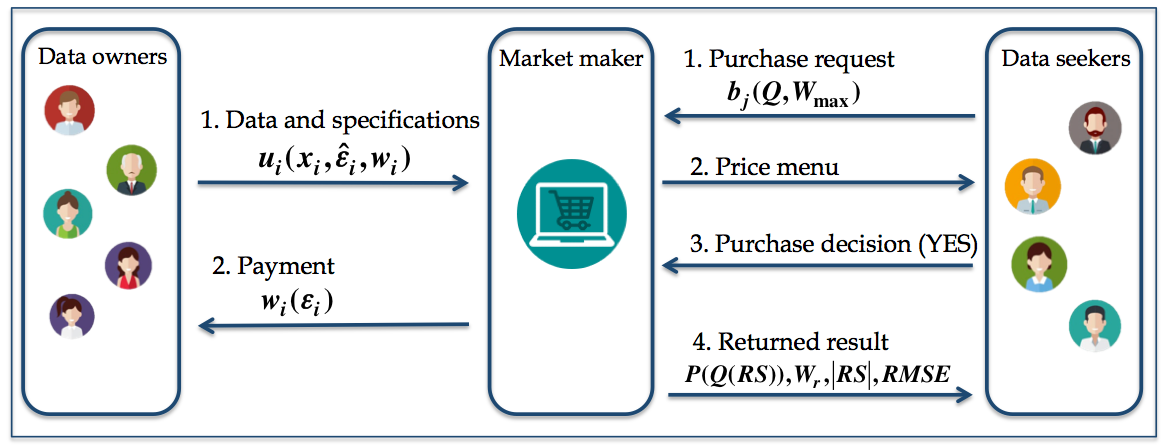}
  \caption{Trading framework for personal data.}
  \label{fig:trading_framework}
\end{figure}

To balance data owners' privacy loss and data buyer's payment to guarantee a fair trade, we propose a personal data trading framework (see Figure \ref{fig:trading_framework}) that involves three main participants: market maker, data owner, and data buyer.

\vspace{-0.65em}

\textbf{Market maker} is a mediator between the data buyer and data owner. Market maker has some coordinating roles. First, market maker serves as a trusted server that answers data buyer's query by accessing the data elements of data owners. Second, a market maker computes and distributes payment to data owners whose data have been accessed while keeping a small cut of the price as a profit $\chi$. Third, a market maker devises some payment schemes for data owners to choose from. Our pricing mechanism is designed to assist the market maker with his or her tasks.
  
\textbf{Data owner} sells his/her data element $x_{i}$ by selecting the maximum tolerable privacy loss $\hat{\varepsilon_{i}}$ and payment scheme $w_{i}$. In DP, $\varepsilon$ is a real non-negative value that is difficult to determine to obtain an exact level of utility. However, \cite{Hsu2014} conducted a study on an economic method of setting $\varepsilon$. Thus, a good user interface is assumed to help data owners understand and determine their $\hat{\varepsilon_{i}}$.

\textbf{Data buyer} purchases an aggregated query answer from the market maker by specifying a query $Q$ and a maximum budget $W_{max}$. Rather than asking the buyer to specify the variance in the query answer, as in \cite{Li2014}, we design our mechanism to be able to obtain the most optimal result with the least noise/errors within the given budget $W_{max}$, since data buyers are highly unlikely to know which value of variance to specify to obtain their desired utility within a limited budget. Thus, designing a mechanism to tackle this issue helps buyers and market maker.

Our framework works as follows. Data owner $u_{i}(x_{i},\hat{\varepsilon_{i}},w_{i})$,  $i\in[1,n]$ sells his/her data element $x_i$ by demanding that the actual privacy loss $\varepsilon_{i}$ must not be greater than their specified $\hat{\varepsilon_{i}}$ while payment should correspond to their selected payment scheme $w_i$. These data elements are stored by a trusted market maker. In the pre-trading stage, the data buyer issues a purchase request by specifying his $Q$ and $W_{max}$. With the request, the market maker will run a simulation and generate a price menu (see Table \ref{tab:pricemenu}) with an average privacy loss $\overline{\varepsilon}$ and a sample size corresponding to prices for the buyer. This price menu provides an overview of the approximate level of utility the buyer may receive for each price.

\vspace{0pt}
\begin{table}[!h]
\setlength{\belowcaptionskip}{-10pt}
\setlength{\abovecaptionskip}{0pt}
  \caption{Example of a price menu.}
  \label{tab:pricemenu}
  \begin{tabular}{|c|c|c|}
	\hline
     Price (\$) & Average privacy loss $\overline{\varepsilon}$ & Sample size\\
	\hline
    5 & 0.039 & 384\\
    50  & 0.545 & 384\\
    100  & 0.619 & 698\\
  \hline
\end{tabular}
\end{table}

The buyer reviews the $\overline{\varepsilon}$ and determines the amount of money he is willing to pay. Once the market maker is notified of the purchase decision, he will run the pricing mechanism (described in Section \ref{sec:pricing_mechanism}) to select a number of representative samples $RS$ from the dataset $x$ and then conduct a query computation by perturbing the answer to ensure the privacy guarantee for all data owners whose data were accessed. Next, the market maker distributes the payment to the data owners in the selected sample $RS$ and returns the perturbed query answer $P(Q(x))$, the remaining budget $W_r$, the size of $RS$, and the root mean squared error $RMSE$ in the query answer. Note that the transaction aborts when the market maker cannot meet their requirements simultaneously.

\section{Pricing Mechanism} \label{sec:pricing_mechanism}
The pricing mechanism directs price and query computations for data buyers and compensation computation for data owners whose data have been accessed. A specially designed pricing mechanism is required in this personal data market because information derived from personal data, unlike other types of physical goods, does not have any tangible properties. Thus, it is difficult to set a price or calculate the traded value as asserted in \cite{sajko2006calculate}. Similarly, \cite{acquisti2013privacy} and \cite{Ruiming2014} discussed why some conventional pricing models (i.e., the cost-based pricing and competition-based pricing models) are not able to price digitalized goods such as data and information. As noted in \cite{shapiro1998versioning}, the only feasible pricing model is the value-based pricing model, through which the price is set based on the value that the buyer perceives. In our framework, the utility of query results determines the price, and this utility is significantly associated with each data owner's level of privacy loss.

\subsection{Baseline Pricing Mechanism}
To simply compute the query price, compensation, and perturbed query result, the baseline pricing mechanism does not involve a sampling procedure. It basically utilizes the entire dataset $x$ in computations to ensure the generation of an unbiased result. In addition, the baseline pricing mechanism implements a simple personalized differentially private mechanism known as the \textit{Minimum mechanism} \cite{Jorgensen2015}, which satisfies $\hat{\varepsilon}_{i}$-PDP by injecting the random noise $X$ drawn from a Laplace distribution with a scale $b$, denoted as $(X\sim Lap(b))$, where $b=1/Min(\hat{\varepsilon}_{1},\hat{\varepsilon}_{2},...,\hat{\varepsilon}_{n})$. The computational run-time of this mechanism is much shorter than that of the sophisticated balanced pricing mechanism, yet it generates a higher query price for a result with more noise. This mechanism does not consider a sophisticated allocation of compensation and perturbation, so it just compensates \textit{all} data owners $u_{i} \in x$ for the same privacy loss $\hat{\varepsilon}_{min}$ and satisfies all $u_{i} \in x$ with the minimum privacy loss $\hat{\varepsilon}_{min}$ resulting in a very low utility (with more noise). For a better result, we propose a balanced pricing mechanism that takes into account the weak performance of the baseline pricing mechanism.

\subsection{Balanced Pricing Mechanism}

In the balanced pricing mechanism, computations are conducted through the use of three main algorithms: \textbf{(1)} sample $h$ subsets of data owners, \textbf{(2)} compute the query price and compensation for all $h$ subsets, and \textbf{(3)} perturb the query answer for all $h$ subsets and then select an optimal subset. 

Algorithm 1 samples $h$ subsets of data owners. It computes the size of an $RS$ representative sample of a dataset $x$ using the statistical method given a dataset $x$, a confidence level score $CLS$, a distribution of the selection $DT$, and a margin of error $MER$. Then, the mechanism randomly selects different/not-duplicated data owners for all the $h$ different subsets. Due to the randomization of data owner selection, the mechanism guarantees an optimal sampling result by increasing the $h$ because an optimal subset $RS$ is selected from all the $h$ different subsets. The output of this algorithm is a set of samples $(RS_1,RS_2,..., RS_h)$ used as an input in Algorithm 2. 

\setlength{\textfloatsep}{10pt}
\vspace{-0.7em}
\begin{algorithm}[]
\SetAlgoLined
\KwData{$x, DT, CLS, MER,$ and $h$}
\KwResult{$(RS_1,RS_2,..., RS_h)$}
 $SS \leftarrow \dfrac{DT*CLS^2}{MER^2}$ \;
 $|RS| \leftarrow \dfrac{SS*|x|}{SS+|x|-1}$\;
 \While{While h$>$0}{
   $RS_h \leftarrow \{u_i|UndupRandomize(1, |x|) | i \in [1, h]\}$\;
   $h \leftarrow h - 1$\;
 } 
 \caption{Sample $h$ subsets of data owners}
\end{algorithm}

\begin{algorithm}[]
\SetAlgoLined
\KwData{$x$, $(RS_1,RS_2,..., RS_h)$, $W_{max}$, $\chi$, $h$, and $\Phi$ }
\KwResult{$W_{p}$, $W_{r}$, and $(\overline{w}_1,\overline{w}_2,..., \overline{w}_h)$}

 $W_{ab} \leftarrow W_{max} - \chi$\;
 \While{While h$>$0}{
 	 $j \leftarrow |RS_h|$\;
     $W_{p} \leftarrow \{\sum_{i=0}^{j} w^{u_{i}\in RS_{h}}(\hat{\varepsilon}^{u_{i}\in RS_{h}}) | i \in [0, j-1]\}$\;

     \eIf{$W_{p} \leq W_{ab}$}{
        \While{While j$<|x|$\&\& $W_{p}< W_{ab}$}{
            $W_{r} \leftarrow W_{ab} - W_{p}$\;
            $RS_h \leftarrow \{u_k|UndupRandomize(1, |x|)\}$\;
            $j \leftarrow j + 1$\;
            \eIf{$W^{r} >  w^{u_{k}\in x}(\hat{\varepsilon}^{u_{k}\in x})$}{
           		$W_{p} \leftarrow W_{p} + w^{u_{k}\in x}(\hat{\varepsilon}^{u_{k}\in x})$\;
                $\varepsilon^{u_{k}\in x} \leftarrow \hat{\varepsilon}^{u_{k}\in x} $\;
     		}{
           		$W_{p} \leftarrow W_{p} + W_{r}$\;
                $w^{u_{k}\in x} \leftarrow W_{r}$\;
                $\varepsilon^{u_{k}\in x} \leftarrow (w^{u_{k}\in x})^{-1}$\;
           	}
     	}
        $W_{r} \leftarrow W_{ab} - W_{p}$\;
     }{
     	$lsTemp \leftarrow RS_{h}$\;
     	$payment \leftarrow 0$\; 
        $W_{r} \leftarrow 0$\;
        $W_{eq} \leftarrow \dfrac{W_{ab}}{|lsTemp|}$\;
		
        \Do{$W_{r} > 0$}{
        	$lsUnderPaid \leftarrow 0 $\;
            \ForEach{$u_{i} \in lsTemp$}{
            	\eIf{$w^{u_{i}}(\hat{\varepsilon}^{u_{i}}) \leq W_{eq}$}{
           			$\varepsilon^{u_{i}} \leftarrow \hat{\varepsilon}^{u_{i}}$\;
                	$payment \leftarrow payment + w^{u_{i}}(\hat{\varepsilon}^{u_{i}})$\;
           		}{
           			$w^{u_{i}} \leftarrow W_{eq}$\;
					$\varepsilon_{i} \leftarrow (w^{u_{i}})^{-1}$\;
                	$lsUnderPaid \leftarrow lsUnderPaid + u_{i}$\;
                	$payment \leftarrow payment + W_{eq}$\;
           		}
            }
            $W_{r} \leftarrow  W_{ab} - payment$\;
        	$W_{eq} \leftarrow W_{eq} + \dfrac{W_{r}}{|lsUnderPaid|}$\;
            $lsTemp \leftarrow lsUnderPaid$\;       
        }

            

		$W_{p} \leftarrow W_{ab}$\;

   }
     
     $\overline{w}_{h} \leftarrow \dfrac{W_{p}}{j} $\;
    
  $h = h - 1$\;
 }
 \caption{Compute query price and compensation for all $h$ subsets}
\end{algorithm}

\begin{algorithm}[]
\SetAlgoLined
\KwData{$x$, $h$, $\Phi$, $(RS_1,RS_2,..., RS_h)$, and $(\overline{w}_1,\overline{w}_2,..., \overline{w}_h)$}
\KwResult{$\overline{\varepsilon}_{max}$, $P(Q(RS)_{opt})$, $\overline{w}_{opt}$, and $RMSE_{opt}$}
	$m \leftarrow h$\;
	\While{While m$>$0}{
    	$\overline{\varepsilon}_{m} \leftarrow \{\dfrac{1}{|RS_{m}|} \sum_{i=0}^{|RS_{m}|} \varepsilon^{u_{i}} | i \in [0, |RS_{m}|-1]\}$\;
	 	$P(Q(RS_{m}))_{k} \leftarrow \{\PE^{Q}(x, RS_{m})*\dfrac{|x|}{|RS_{m}|} |k \in [0, \Phi-1]\} $\;
     
     	$RMSE_{m} \leftarrow \bigg \lbrace \sqrt{\dfrac{\sum_{k=0}^{\Phi}(P(Q(x) - P(Q(RS_{m}))_{k}))^2 }{\Phi}} | k \in [1, \Phi-1] \bigg \rbrace$\;
		$m = m - 1$\;
 	}
    $\overline{\varepsilon}_{max} \leftarrow Max(\overline{\varepsilon}_1,\overline{\varepsilon}_2,..., \overline{\varepsilon}_h)$\;
    
    $optIndex \leftarrow \{index | \overline{\varepsilon}_{index} = \overline{\varepsilon}_{max}\}$\;

    $RMSE_{opt} \leftarrow RMSE_{optIndex}$\;
    $P(Q(RS))_{opt} \leftarrow P(Q(RS))_{optIndex}$\;
    $\overline{w}_{opt} \leftarrow \overline{w}_{optIndex}$\;

\caption{Perturb the query answer for all $h$ subsets and then select an optimal subset}
\end{algorithm}

Algorithm 2 computes the query price and compensation for all the $h$ subsets. Given a data buyer's maximum budget $W_{max}$, query $Q$, dataset $x$, number of samples $h$, number of perturbations $\Phi$, market maker's benefit $\chi$, and $h$ subsets $(RS_{1},RS_{2},...,RS_{h})$ from Algorithm 1, the algorithm returns the query price $W_{p}$, remaining budget $W_{r}$ (if applicable), compensation $w_{i}(\varepsilon_{i})$ for each $u_{i}$, and average compensation $\overline{w}_{i}$ for each subset because Algorithm 3 uses this result to select an optimal subset from all $h$ subsets. The algorithm first computes the available budget $w_{ab}$ by subtracting $\chi$ from the given $W_{max}$. Next, the algorithm computes the total payment $W_{p}$ required when paying for the maximum privacy loss $\hat{\varepsilon}_{i}$ of $u_{i} \in RS_{h}$. $w^{u_{i}\in RS_{h}}(\hat{\varepsilon}^{u_{i}\in RS_{h}})$ denotes a payment for data owner $u_{i}$ in $RS_{h}$ for $\hat{\varepsilon}_{i}$. When $W_{p}$ is smaller than $W_{ab}$, the algorithm pays each $u_{i}$ for $\hat{\varepsilon}_{i}$ while using $W_{r}$ to include more data owners into $RS_{h}$ by paying for $\hat{\varepsilon}_{i}$ or $\varepsilon_{i}<\hat{\varepsilon}_{i}$ based on $W_{r}$. This process repeats until all $W_{r} = 0$ or $|RS_{h}| = |x|$, as the utility is influenced by both the size of $RS$ and by the privacy loss $\varepsilon_{i}$ of all $u_{i}$. Otherwise, when $W_{p}>W_{ab}$, the algorithm determines the equal payment $W_{eq}$ for each $u_{i} \in RS_{h}$ and then verifies if each $u_{i}$ should be paid exactly $W_{eq}$ or less when $w^{u_{i}}(\hat{\varepsilon}^{u_{i}}) < W_{eq}$. The updated $(RS_{1},RS_{2},...,RS_{h})$ as an output is used in Algorithm 3.

With the output of Algorithm 2, Algorithm 3 perturbs the query answer and selects an optimal subset from all $h$ subsets. It computes the average privacy loss $\overline{\varepsilon}$ and perturbed query result $P(Q(RS))$ based on the proportional difference between $x$ and $RS_h$ by multiplying result of $\PE$ by $|x|/|RS_h|$, and $RMSE$ in each $(RS_{1},RS_{2},...,RS_{h})$. It then selects an optimal $RS$ with a maximum average privacy loss of $\overline{\varepsilon}_{max}$ denoting a high probability that less random noise is included in the result. Finally, the algorithm finds the corresponding $RMSE_{opt}$, $P(Q(RS))_{opt}$ and $\overline{w}_{opt}$ of the optimal $RS$ selected.   

At the end, data buyers receive the perturbed query answer $P(Q(RS))$ along with the remaining budget $W_{r}$ (when applicable), the number of data owners in $RS$, and the mean squared error $RMSE$ in the query answer. Data owners are then compensated according to their actual privacy losses $\varepsilon_{i}$.

\section{Experiment}\label{sec:evaluation}

\textbf{Experimental setup:} We divide the experiment into two components: (1) the simulation of our balanced pricing mechanism and (2) the comparison of our mechanism with the baseline pricing mechanism. We examine the query price $W_{p}$, root mean squared error $RMSE$, average privacy loss $\overline{\varepsilon}$ and average compensation $\overline{w}$ that each $u_{i}$ obtained from both mechanisms and then conclude that for the same $W_{p}$, which mechanism generates the smallest $RMSE$ value. Due to space constraints, we only show the experimental result of the following count query \textit{Q: "How many people commute by personal car in the USA?"}

\textbf{Data preparation:} From our survey, we obtained 486 records of personal data from 486 data owners. To generate more accurate experimental results, a larger dataset is preferable, so we duplicated our survey dataset 500 times to obtain a larger dataset $x$ of 243,000 records. To conduct such an experiment, each data record must have two important variables: the maximum tolerable privacy loss $\hat{\varepsilon}$ and a payment scheme $w$. For the sake of simplicity, we assume $\hat{\varepsilon}\in[0, 1]$ and two types of payment schemes (as described in Section \ref{sec:principle_truthful_privacy_valuation}). In preparing our data, we generate these two variables for each record/data owner according to the survey answers. When a data owner has chosen to have \textit{very high} and \textit{high} alterations/perturbations, they are classified under the \textit{conservative} group, so his or her $\hat{\varepsilon_{i}}$ values are set to 0.1 and 0.3, respectively. For \textit{low} and \textit{very low} perturbations, the $\hat{\varepsilon_{i}}$ values are set to 0.7 and 0.9 respectively, and such data owners are categorized under the \textit{liberal} group. To optimize their benefits, we set the most optimal payment scheme for them based on their $\hat{\varepsilon_{i}}$ values. For the conservative group with $\hat{\varepsilon_{i}}$ values of 0.1 and 0.3, we set a payment scheme \textit{type A}, while \textit{type B} is set for liberal group with $\hat{\varepsilon_{i}}$ values of 0.7 and 0.9. In turn, we obtain a usable and large dataset for our experiment.

\begin{figure*}[]
\centering
\setlength{\belowcaptionskip}{-13pt}
\setlength{\abovecaptionskip}{0pt}
\mbox{\subfigure[Query price and RMSE.]{\includegraphics[width=2.3in]{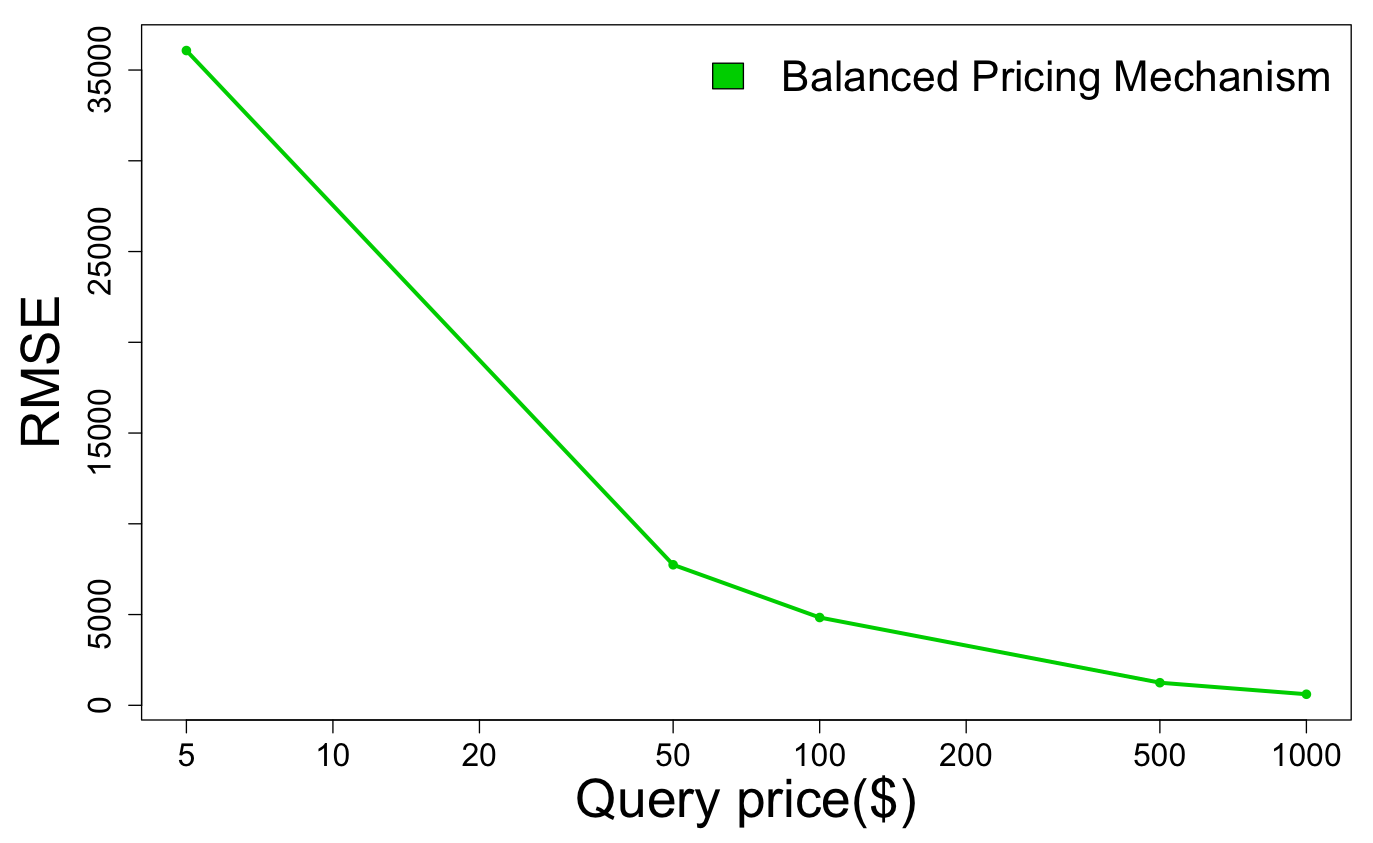}}\quad
\subfigure[Query price and $\overline{\varepsilon}$.]{\includegraphics[width=2.3in]{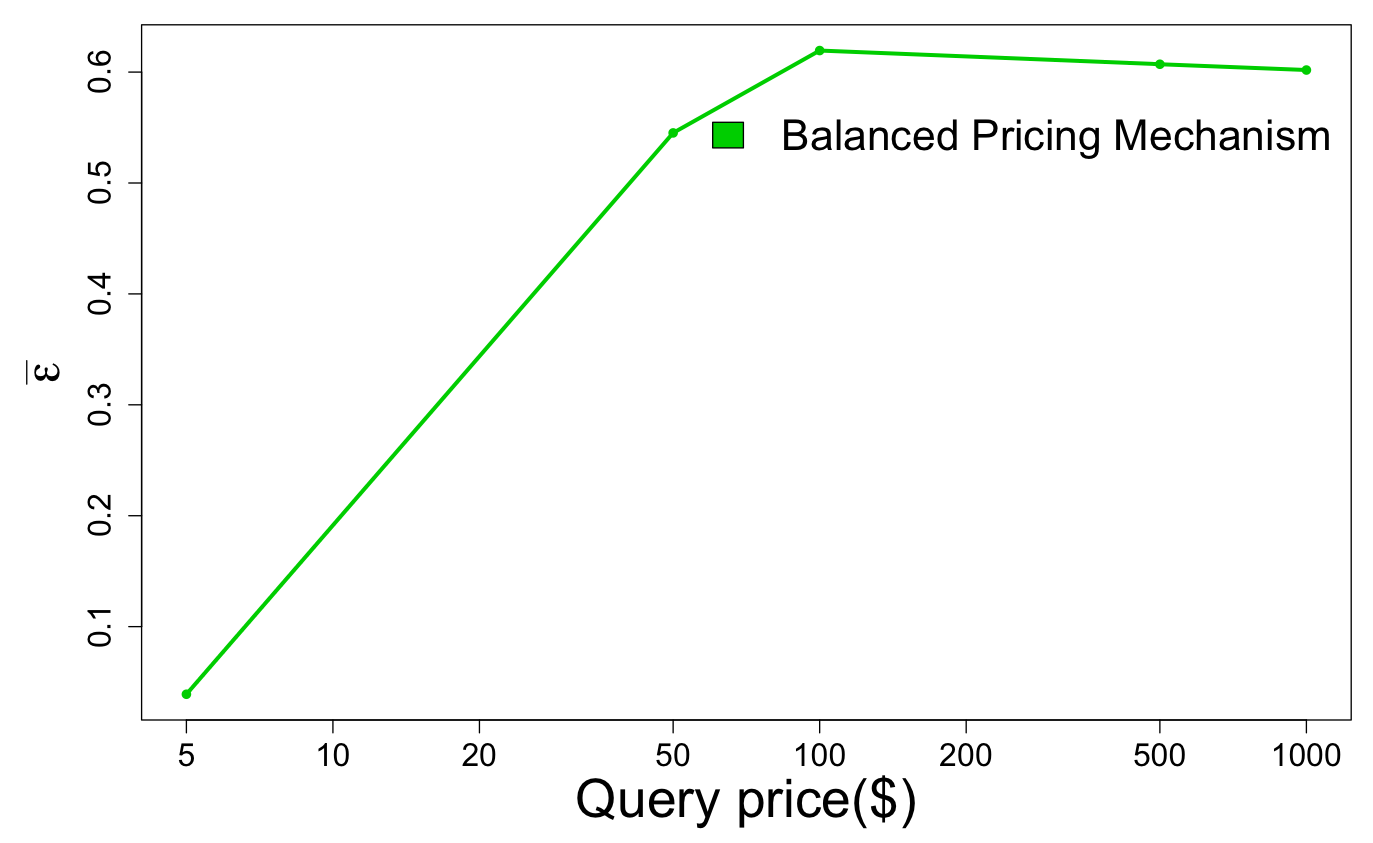}}\quad
\subfigure[Query price and average compensation.]{\includegraphics[width=2.3in]{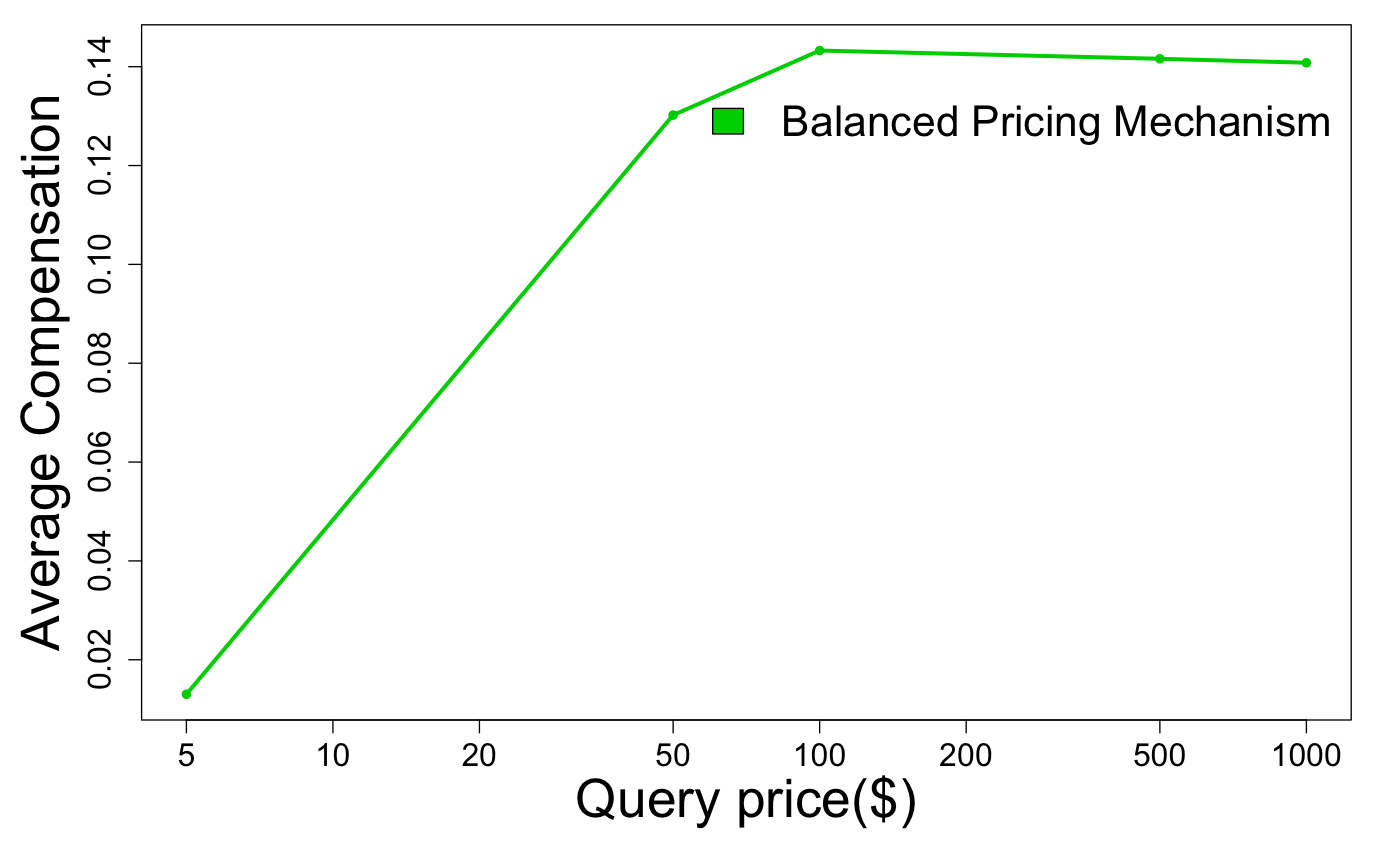}}
}
\caption{Simulation on \textit{balanced pricing mechanism}.} \label{fig:simulation}
\end{figure*}

\begin{figure*}[]
\centering
\setlength{\belowcaptionskip}{-8pt}
\setlength{\abovecaptionskip}{0pt}
\mbox{\subfigure[Query price and RMSE of both mechanisms.]{\includegraphics[width=2.3in]{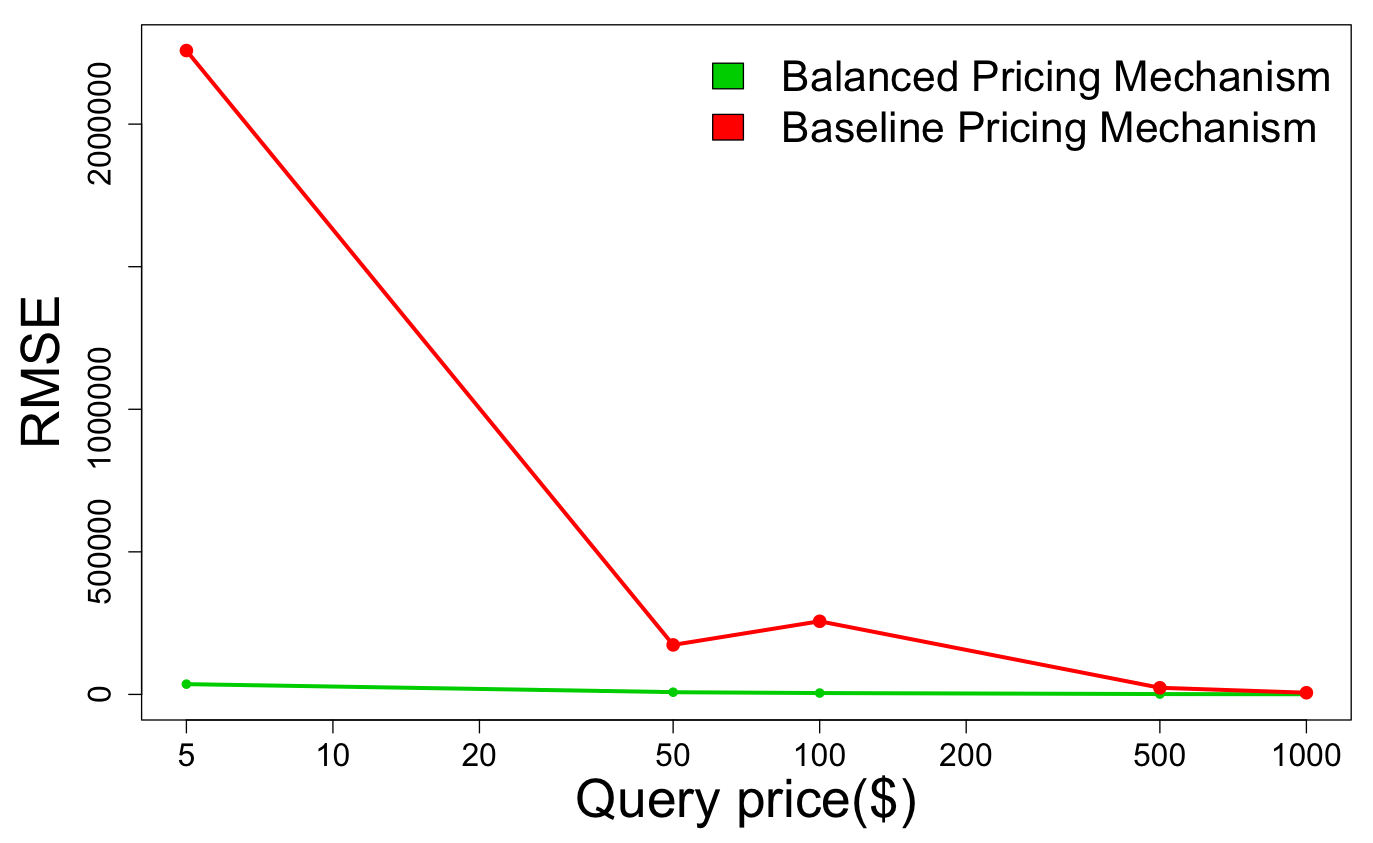}}\quad
\subfigure[Query price and $\overline{\varepsilon}$ of both mechanisms.]{\includegraphics[width=2.3in]{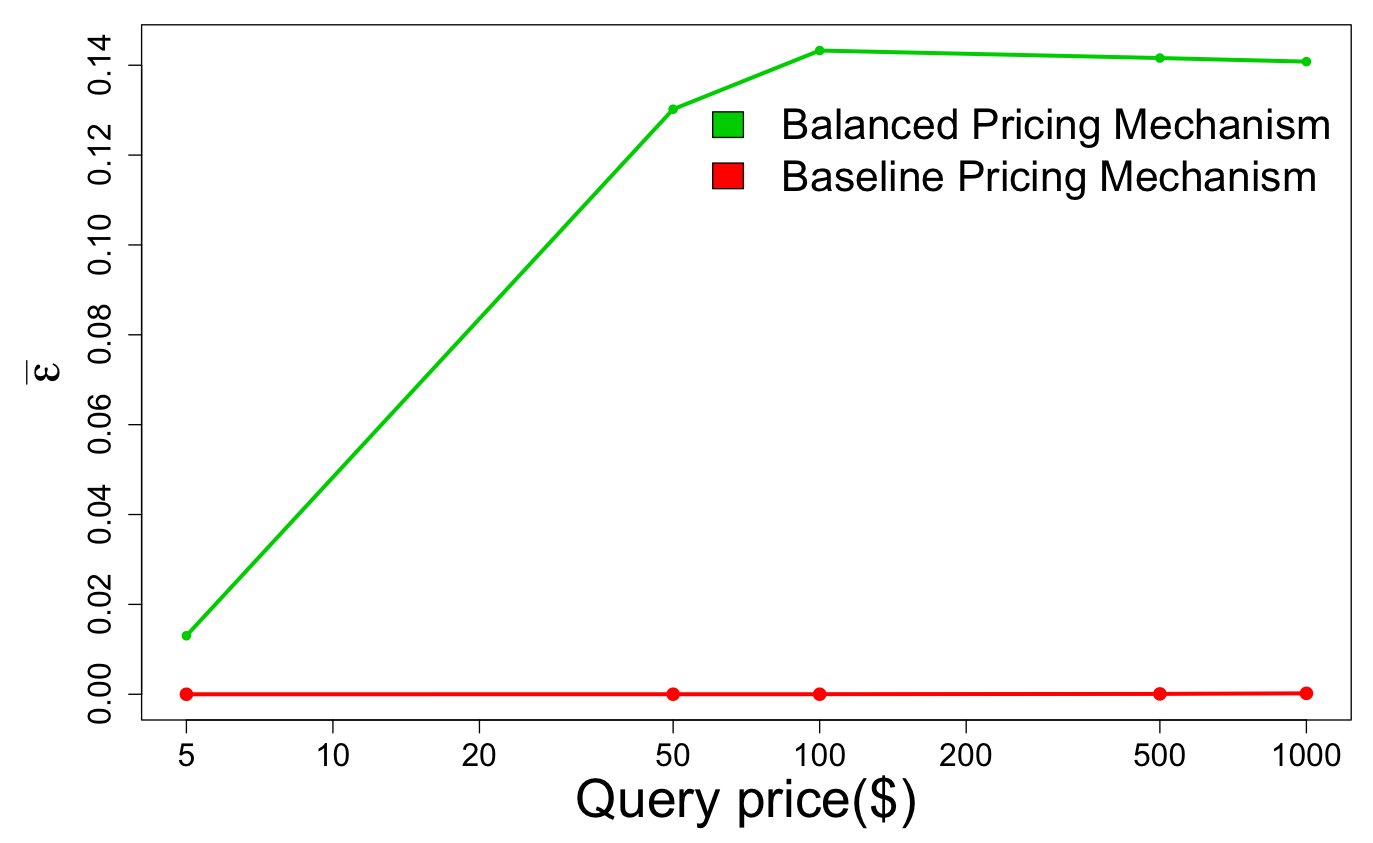}}\quad
\subfigure[Query price and average compensation of both mechanisms.]{\includegraphics[width=2.3in]{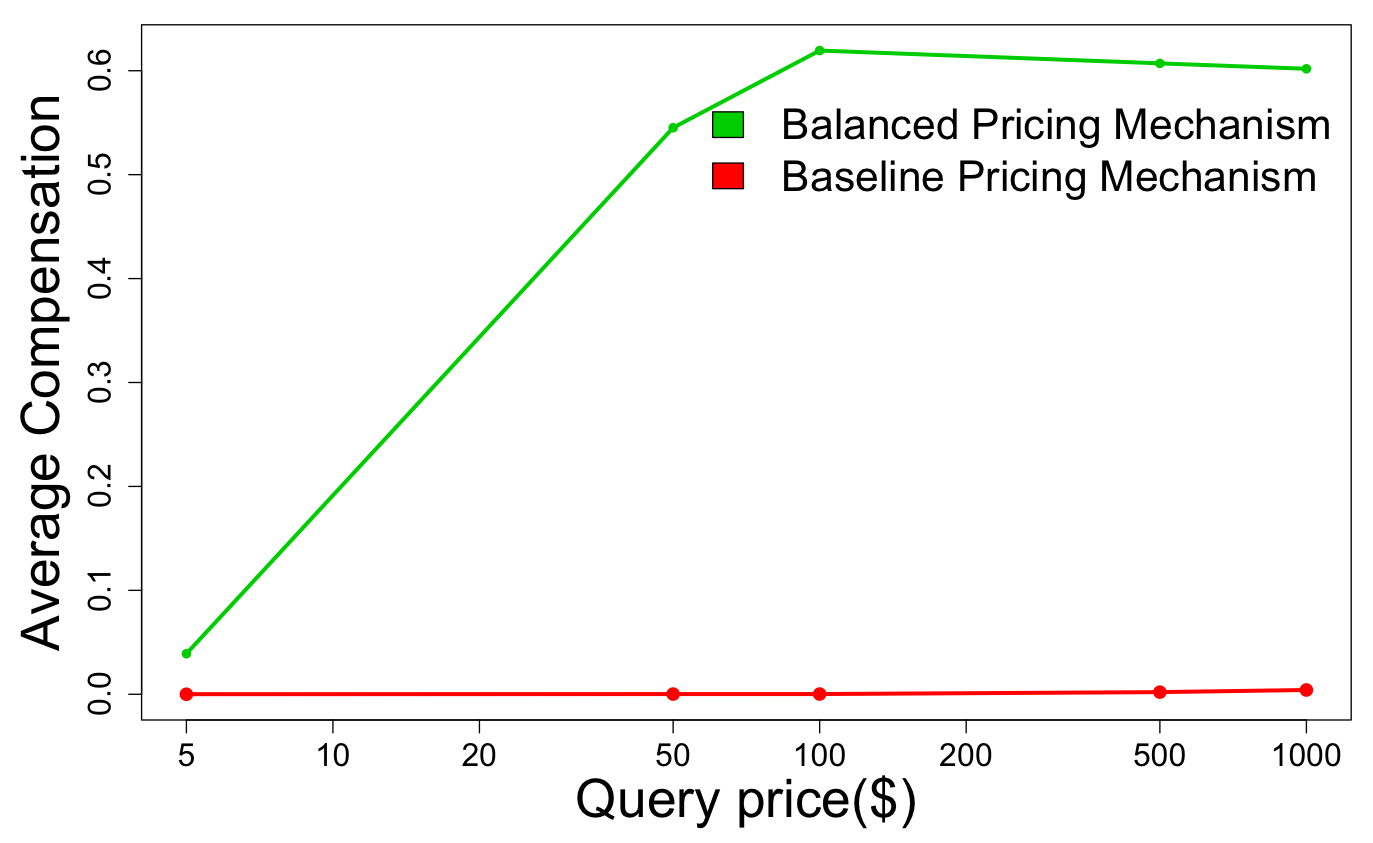}}
}
\caption{Comparison between the \textit{balanced pricing mechanism} and \textit{baseline pricing mechanism}.} \label{fig:comparison}
\end{figure*}

\textbf{Experiment results:} We first conduct a simulation of our mechanism (Figure \ref{fig:simulation}) to explain the correlation between the query price and RMSE , between the query price and average privacy loss $\overline{\varepsilon}$, and between the query price and average compensation value. Figure \ref{fig:simulation}a shows that the $RMSE$ value decreases as the query price increases. This pattern is reasonable in practice because the higher the query price is, the lower the $RMSE$ should be. Remarkably, the $RMSE$ value declines dramatically with query price from \$5 to \$50 but then gradually and slightly decreases for \$50 to \$1000. We can attribute this phenomenon to the impact of privacy parameter $\varepsilon_i$ of each data owner $u_i$ and to the number $|RS|$ of data owners responding to the query. When the query price is approximately \$50 or less, it can only cover the compensation of $RS$, so with the same size $|RS|$, an increase in the query price (i.e., \$5 to \$50) can also increase the $\overline{\varepsilon}$ value in $RS$. However, when the query price exceeds what is needed to pay for $\hat{\varepsilon}_i$ for all $u_i$ in $RS$, the remaining budget is used to include more data owners in $RS$, which can significantly or marginally decrease the overall $RMSE$ while increasing the $\overline{\varepsilon}$ value depending on the distribution of data. When more conservative data owners are included in $RS$, this can affect the $\overline{\varepsilon}$ value resulting in just a minor decrease in $RMSE$ despite more money being spent. For this reason, the price menu plays a crucial role in providing an overview on approximate degree of change in $RMSE$ values corresponding to query prices. In turn, data buyers can decide whether it is worth spending more money for a minor decrease of $RMSE$ within their available budgets. Figure \ref{fig:simulation}b and Figure \ref{fig:simulation}c show a similar correlation pattern between the query price and $\overline{\varepsilon}$ and between the query price and $\overline{w}$. They show that the higher the query price is, the higher $\overline{\varepsilon}$ and $\overline{w}$ values become. A marginal decrease in $\overline{\varepsilon}$ with a significant rise in query price (\$100 to \$1000) shown in Figure \ref{fig:simulation}b can be attributed to the phenomenon illustrated in Figure \ref{fig:simulation}a whereby the $RMSE$ value only slightly decreases within a significant price increase.

We next compare the results of our \textit{balanced pricing mechanism} with those of the \textit{baseline pricing mechanism} (Figure \ref{fig:comparison}). The experimental results show that our balanced pricing mechanism considerably outperforms the baseline mechanism under almost all conditions. Figure \ref{fig:comparison}a shows that our balanced mechanism produced a noticeably smaller $RMSE$ value for the same query price relative to the baseline mechanism. In particular, our balanced mechanism produced a significantly smaller $RMSE$ value even when the query price was set to be relatively low (i.e., \$5) because instead of querying from the entire dataset, our balanced pricing mechanism only queries from a representative sample $RS$. This reduces the query price while still generating a smaller $RMSE$. Due to random noise drawn from the Laplace distribution, we can see that the $RMSE$ of the baseline mechanism, rather than declining, rises for query prices \$50 to \$100. Figure \ref{fig:comparison}b and Figure \ref{fig:comparison}c show a similar pattern in that the $\overline{\varepsilon}$ and $\overline{w}$ of our balanced pricing mechanism are significantly higher than those of the baseline mechanism.

\section{Discussion}

The above listed experiment results show that our balanced pricing mechanism considerably outperforms the baseline pricing mechanism. This is attributed to two main factors. First, we apply an \textit{exponential-like $\PE$ mechanism} (see Definition \ref{def:PE_mechanism}) to achieve \textit{Personalized Differential Privacy (PDP)} to take advantage of the individual privacy parameter $\hat{\varepsilon}$ of data owners, especially of the liberal group. In contrast, the baseline mechanism can only apply a \textit{minimum mechanism} to achieve \textit{PDP} by adding a large amount of random noise drawn from Laplace distributions utilizing the smallest $\hat{\varepsilon}$ of the entire dataset. Second, our mechanism produces a considerably smaller $RMSE$ for the same query price. In other words, for the same level of utility, we can indeed reduce the query price, as our mechanism only queries a small subset of a dataset while generating unbiased results from a random sampling and selection procedure. we thus exclusively compensate the data owners of the queried subset, while the baseline mechanism must compensate all data owners of a dataset to run a query on the dataset to obtain unbiased results. Therefore, our balanced pricing mechanism is more efficient than the baseline mechanism.   

In the price menu, it is important to illustrate trends of higher prices and higher levels of approximate utility (denoted as $\overline{\varepsilon}$). However, Figure \ref{fig:simulation}b shows a slight decrease in $\overline{\varepsilon}$ from \$100 to \$1000. This phenomenon could be attributed to the number of samplings $h$ applied in the mechanism. Despite showing a budget increase, it cannot fully guarantee that $\overline{\varepsilon}$ will increase due to the random selection of data owners with various $\hat{\varepsilon}$ values. Thus, our naive solution is to increase the price gap in the price menu to guarantee a distinguished increase in $\overline{\varepsilon}$ for an increasing query price. More discussion on this point will be included in our next work.

It is also crucial to ensure that data owners can technically choose an appropriate maximum tolerable privacy loss $\hat{\varepsilon}_{i}$ that reflects their privacy attitude and risk orientation. This problem indeed remains an open question in the differential privacy community regarding how to set the value of $\varepsilon$ or $\hat{\varepsilon}$ in our setting. Although \cite{Hsu2014} proposed an economic method for choosing $\varepsilon$, this problem has not been widely discussed. A part of solution, we provide some options of $\hat{\varepsilon}=\{0.1, 0.3, 0.7, 0.9\}$ corresponding with \{very high, high, low, very low\} data perturbation level. Very high perturbation (i.e., $\hat{\varepsilon}=0.1$) means that more random noise is added to the result, so the data owners have a very high privacy guarantee. However, some data owners might not understand how the perturbation works, so we can provide an interactive interface allowing them to see the approximate change on their actual data for a different value of $\hat{\varepsilon}$. A similar concept of the interactive interface\footnote{http://content.research.neustar.biz/blog/differential-privacy/WhiteQuery.html} is used to explain a perturbation via Laplace mechanism. Thus, we can create a similar interface for exponential-like data perturbation mechanism to assist data owners and buyers to understand the meaning of $\hat{\varepsilon}$.

\section{Related Work}\label{sec:related_work}

In the field of pricing mechanism design, there are two crucial focuses of research: \textit{auction-based pricing} and \textit{query-based pricing}. Auction-based pricing has attracted the attention of \cite{fleischer2012approximately}, \cite{Ghosh2015}, \cite{Riederer2011}, and \cite{Roth2012}. Auction-based pricing allows data owners to report their data valuations and data buyers to place a bid. From a practical point of view, it is very difficult for individuals to articulate their data valuations as reported in \cite{acquisti2013privacy}. Moreover, the price described in \cite{Riederer2011} is eventually determined by the data buyer without considering data owners' privacy valuations or actual privacy losses. On the other hand, query-based pricing, as defined in \cite{Koutris2012}, involves the capacity to automatically derive the prices of queries from given data valuations. The author in \cite{Koutris2012} also proposes a flexible arbitrage-free query-based pricing model that assigns prices to arbitrary queries based on the pre-defined prices of view. Despite this flexibility, the price is non-negotiable. The buyer can obtain a query answer only when he or she is willing to pay full price. Unfortunately, this model is not applicable to personal data trading, as it takes no account of issues of privacy preservation. \cite{Li2014} extended and adapted the model by applying differential privacy for privacy preservation and for the quantification of data owners' privacy losses, yet this method still presents a number of problems, as explained in Section \ref{subsec:principles}.

\section{Conclusion and Future Work}\label{sec:conclusion_future_work}
We analyzed people's privacy attitude and levels of interest in data trading, then identified five key principles for designing a reasonable personal data trading framework. For an operational market, we proposed a reasonable personal data trading framework based on our key principles. In addition, we proposed a balanced pricing mechanism that balances money with privacy to offer more utility to both data owners and data buyers without circumvention. Finally, we conducted various experiments to simulate our mechanism and to prove its considerably higher degree of efficiency in comparison to a baseline pricing mechanism. The results show that our study has identified and tackled some radical challenges facing the market, thus facilitating the existence of the personal data market.

Having investigated the challenges of this market, we identify a number of interesting avenues for future work. To obtain an optimal query answer and price, it is crucial to carefully design a payment scheme using game theory. In the present study, we only designed two types of payment schemes for liberal and conservative data owners. We will develop a more sophisticated design in our future work. Moreover, in our study, a market maker is assumed to be a trusted server storing and accessing data owners' data on their behalf, yet to some extent, trust has become a difficult question to address from both technical and social standpoints. Thus, for future work, we can consider a trading framework and pricing mechanisms in which market makers are assumed to be untrustworthy..

\begin{acks}
This work was supported through the A. Advanced Research Networks JSPS Core-to-Core Program. The work was also supported through JSPS KAKENHI Grant Numbers 16K12437 and 17H06099.

\end{acks}